\documentclass{ws-tech}
\usepackage{graphics}
\usepackage{amssymb}
\usepackage{amssymb,amsmath}
\usepackage[T1]{fontenc}
\usepackage{setspace}
\usepackage{textcomp}
\usepackage{dsfont}
\usepackage{mathrsfs}
\usepackage{wrapfig}
\usepackage{multirow}
\usepackage{multicol}
\usepackage[dvipsnames,svgnames,table]{xcolor}
\usepackage{ulem}
\DeclareMathAlphabet{\mathpzc}{OT1}{pzc}{m}{it}

\begin{document}
\markboth{Nilo Sylvio Costa Serpa}
{}
%
\catchline{}{}{}{}{}
%

\title{Nilo Sylvio Costa Serpa, $\mathpzc{MSc}$ - Astronomy}
\maketitle
\vspace{-30mm}
\begin{center}
{\it Instituto de Ci\^{e}ncias Exatas e Tecnologia\\
UNIP - Universidade Paulista\\
SGAS Quadra 913, Conj B\\
Bras\'{i}lia - DF, Brasil\\}
\end{center}
\vspace{4.0mm}
\begin{figure}[h]
\begin{center}
\includegraphics[scale=0.64]{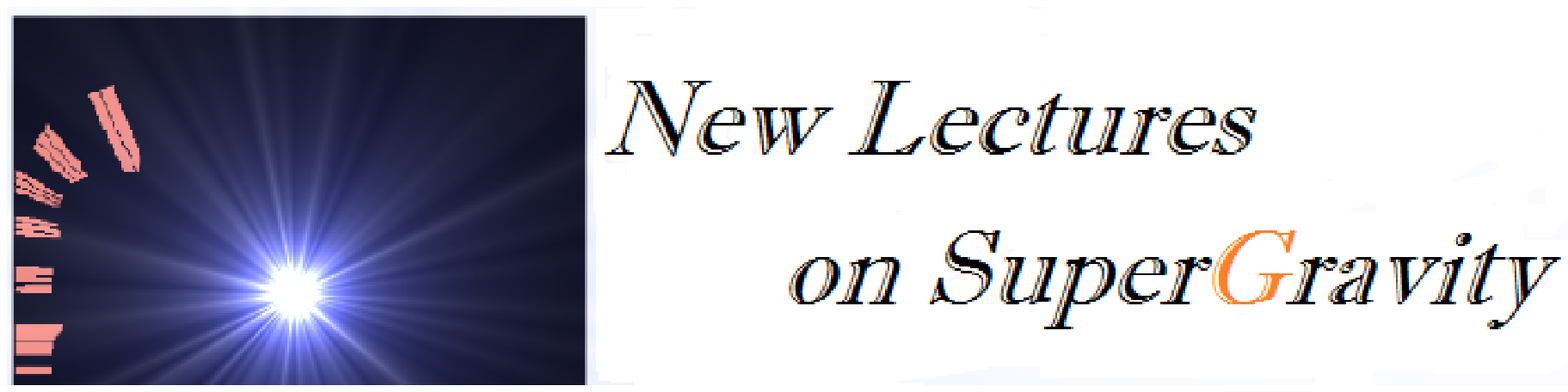} 
\end{center}
\end{figure}

\author{{\bf Abstract}}
\\
\begin{abstract}
This essay aims to summarize the main physical features arising from a new supersymmetric theory of gravitation.  Based on preliminary discussions about classical field theory, cosmology, algebra and group theory, and taking formal results and theoretical considerations in comparison with several contributions from great authors, present work deals with gravity inside the limits of a meta-field theory, that is, a non-quantized but consistent representation of supergravity, the supersymmetry between gravitons and gravitinos. The introduction of meta-fields furnishes an independent framework for the study of gravity despite of constraints of quantization, treating the supersymmetric partners as deterministic actors of gravitation and not simply probabilistic entities. I explain my belief that gravitational field, by its own nature, is not quantizable in the same foot as the other fields, what does not means that we can not understand gravity by similar formal veins. Also, present work proposes the implementation of the so-called {\bf SCYL} program (Supersymmetric Cosmology at Yonder Locals), an attempt to apply supergravity as a meta-field theory to solve problems on astrophysical cosmology. The SCYL program is an effort directed to search convergence between the knowledge brought by cosmology and by supersymmetry to attain more clearness about the structure and evolution of the Universe. As an additional motivation, there are some exercises in the final part of the work to aid fixing of the main concepts.\\       
\end{abstract}


\vspace{36in}
$$
$$
\hrule
\begin{flushright}
{\it Supersymmetry is a beautiful symmetry between bosons and fermions, 
although there is no evidence of it in Nature. This does not mean 
that it is not present, but that it must be well hidden}.
\\
\vspace{3.2mm}
$\mathpzc{Pierre \hspace{0.4mm} Fayet}$
\end{flushright}
\newpage
\pagestyle{plain}
\tabela
\newpage
\pagestyle{headings}
\pagenumbering{arabic}
\renewcommand*\thesubsection{\arabic{subsection}}
\setcounter{page}{1}

\section*{Acknowledgements}
\addcontentsline{toc}{section}{Acknowledgments}
 I thank Dr. José-Abdala Helayel Neto for the stimulating support. Special thanks to Dr. Camilo Tello at INPE - Instituto Nacional de Pesquisas Espaciais - whom contributed with some suggestions. I've benefited from conversations with many colleagues, including Julio Peçanha, Marcelo Byrro and Antonio Teixeira. I also would like to thank the encouragement and patience of my family. 
$$
$$
\begin{center}
$\diamondsuit\diamondsuit\diamondsuit$
\end{center}

\newpage

\section*{Prologue} 
\addcontentsline{toc}{section}{Prologue}

This essay is a compilation of my ultimate work on the former model for supergravity I proposed at 2002, now reaching its final theoretical outcomes and the top point with the so-called SCYL program (Supersymmetric Cosmology at Yonder Locals). More ontological than phenomenological, present essay also includes some particular notes I made during several classes on supersymmetry and supergravity cursed at CBPF - Centro Brasileiro de Pesquisas Físicas - and from talks with Professors Bert Schroer, Wolfgang Bietenholz and José-Abdala Helayel Neto. Fragments of my lectures on gravitation, wrote before many major advances in the theory, are revised, adapted and resumed to give more clearness for readers that are not entirely familiarized with the principal publication. This study treats mainly the Abelian side of the theory (non-Abelian trend is being taken for another essay). It is not for one that desires to learn field theory and it must be read like a book; it is necessary to be familiarized with basic formalism. I claim attention for the fact that this original theory tries to indicate the true role of gravitino field in the supersymmetric theory of gravity. In particular, the last chapter is my project (the SCYL program) to be sponsored by TechSolarium - Solar Technology -, based in great on the previous chapters. 
$$
$$
\begin{center}
$\diamondsuit\diamondsuit\diamondsuit$
\end{center}

\newpage

\section*{Conventions and basic definitions} 
\addcontentsline{toc}{section}{Conventions and basic definitions}

 In order to aid the reader, I did here a resume of the principal notations. 
\begin{itemize}
\item In the SCYL program I called "cold" the relic gravitinos from early stages of the Universe, and "hot" the gravitinos generated by dense massive objects. 
\item Any letter with an overdot ($\dot R$, $\dot \theta$, etc.) denotes time derivative of the quantity that it represents.
\item Any letter overwrote by a right coma (${R'}$, ${\theta'}$, etc.) denotes positional derivative of the quantity that it represents.
\end{itemize}
$$
$$
\begin{center}
$\diamondsuit\diamondsuit\diamondsuit$
\end{center}

\newpage

\section*{Preliminaries} 
\addcontentsline{toc}{section}{Preliminaries}
$$
$$
\hrule
\begin{flushright} 
\small {\it Symmetries in modern physics have taken an even
stronger role to such an extent that the laws of modern
physics cannot even be formulated without the concept
of symmetries. To make the framework of local
quantum field theory meaningful, symmetries have to
be invoked from the very beginning.}
\\
\vspace{3.2mm}
$\mathpzc{Julius \hspace{0.4mm} Wess}$
\end{flushright}

Supersymmetry (SUSY) is a Bose-Fermi symmetry referring to the spectrum of coupling energy among particles; it is a device that tries to fulfill a phenomenological gap between the sectors of spectrum related to electroweak interactions and GUT scale (from $10^{2}$ Gev to $10^{16}$ Gev). The gap results from the second Higgs quantization, needed in Weinberg-Salam, forcing the introduction of SUSY mechanisms to provide intermediary physics inside those limits. Successive symmetry breaks are in part supplied by gravitic fields that do not couple (at least in thesis) with matter. For a more complete treatment and the necessary field theory background, please, see the standard literature at the end of this work. Supergravity (SUGRA) is the supersymmetry that occurs in gravity. The smallest theory of supergravity relates two types of fields referring to the hypothetical particles graviton and gravitino. The relevance of supergravity to cosmology is that it offers an effective field theory behind the expanding universe and timedependent scalar fields. In particular, as we shall see, the consideration of timedependence as a fundamental feature of the fields in the model I proposed few years ago defines supergravity itself as a theory that describes gravity by cumulative effects on matter.  

Gravitorial theory, a sort of supergravity theory, was born to describe supersymmetry in gravity as an outcome of long-time cumulative effects of gravitation. It settles supergravity in a formal "forequantum" \hspace{0.2 mm} representation as a non-local field theory with local vestiges of the time flowing process, which culminates in scenarios of supersymmetry spontaneous break. We may say it is a "neo-classic"\hspace{1.0 mm} sight of supergravity, since it does not presuppose probabilistic interpretation of the formalism, either operatorial devices previous and irremediably linked to observers. Even so, theoretical unrolling seems familiar, as gravitors and spinors are similar affinor structures.

Quantum field theories are usually built by applying quantization rules to a continuum field theory; this demands the replacement of Poisson brackets by commutators or anticommutators, as we deal with bosonic or fermionic fields respectively. In fact, gravitorial theory is more concerned to supergravity itself, and not to conjectures about the quantum nature of space and time, in the same critic sense pointed by Butterfield and Ishami (2000):
\\
\\
"...general relativity is not just a theory of the gravitational field - in an appropriate sense, it is also a theory of spacetime itself; and hence a theory of quantum gravity must have something to say about the quantum nature of space and time. But though the phrase 'the quantum nature of space and time'\hspace{0.8 mm} is portentous, it is also very obscure, and opens up a Pandora's box of challenging notions".
\\
\\
On some sort of theory of quantum gravity they observe that "...it is wrong to try to construct this theory by quantizing the gravitational field, i.e., by applying a quantization algorithm to general relativity (or to any other classical theory of gravity). We shall develop this distinction between the general idea of a theory of quantum gravity, and the more specific idea of quantized version of general relativity...". 

Perhaps one may consider a "quantum of spacetime"\hspace{0.8 mm} as in the times of Zenon of Elea, when Proclo understood a given number as a cutoff on the "compact thickness"\hspace{0.8 mm} of the continuum of the real numbers; a quantum of spacetime would be a type of occasional physical "cutoff"\hspace{0.8 mm} in the continuum of the cosmic tapestry. However there is hope that supergravity is a consistent quantum theory of gravity, I believe in a theory essentially "classic"; my concern is to specify a representative frame of the involved symmetries, preceding to statistics or stochastic behavior due to experimental activity, and, thereby, more realistic in the sense of a description of the world that foregoes operatorial interventions (of course we may think, focusing gravitational waves, in "amplitude operators"\hspace{0.8 mm} within experimental perspectives, as textbooks customarily use to replace momentum $p$ by the operator $i\hbar \nabla$ to compute $\left| {\langle b\left| {H\left| {a\rangle } \right.} \right.} \right|$, but this is not yet the case considering the present technology and the fact that the virtual carrier of gravity, the graviton associated to the gravitational wave, is not acted upon by the strong and the electromagnetic interactions). Quantization of gravitorial theory is a matter of lab procedures, and even so there is no certainty whether we are correct to trying to quantize gravity, as exposed above, and whether it would be possible to detect supergravity partners in laboratory.
\subsection*{Why symmetry is important?} 
\addcontentsline{toc}{subsection}{Why symmetry is important?}
The Greeks developed the symmetry concept. To them, symmetry meant proportionality, and, in especial, commensurability, an adequate Latin mirror translation for the original Greek meaning of "syn-metry". So, it is clear that in early times symmetry had spatial meaning. Only when symmetry became a more abstract concept it appeared that invariance through time is part of the symmetry concept too. In fact, symmetry is important because human beings always try to identify regular patterns looking for the best understanding of natural processes. The human brain works in this way, and, although sometimes it may lead to mistakes and erroneous impressions, nature seems to evolve by symmetries or, at least, to induce human brain to see the world as it was made by symmetries. The growing of crystals, the shape of the vertebrates, the snow flocks, the Gaussian distributions applicable to several facts, are good examples of the general tendency to find symmetric patterns. Of course, there are shapes more and more complicated in nature, then with weaker symmetry. But, as everybody, including physicists, has his own idea of what is symmetry, I mean it in the representational framework of group theory.

The notion of symmetry plays a great role in quantum physics. In modern quantum field theories, symmetries are important because we might get information about the system without any knowledge on the real laws behind its dynamics (Wess, 2009). In fact, we start by the implementation of the symmetries themselves to be able to enunciate formally the laws. This way to think a dynamical system is now so deeply ingrain in our minds that we say to understand it if we find an aesthetic symmetry in such system. 
Supersymmetry is an uncommon deep generalization of symmetry. This generalization was attained by the fact that not only commutators, which the symmetries can be formulated, but also anticommutators are very useful, especially when we deal with particles with half integer spin. Thus, the idea is to formulate a general concept of symmetry, the supersymmetry, in terms of commutators and anticommutators as well. Although there is no experimental
confirmation of supersymmetry in nature, it has influenced greatly the theoretical work in high energy physics.   
\subsection*{What does quantum mechanics say?} 
\addcontentsline{toc}{subsection}{What does quantum mechanics say?}
The concept of physical probability was really born with the adventure of quantum mechanics, even though in the core of this discipline it has been treated systematically as the expression of the inexact knowledge.
The focal point was to interpret the so called wave function $\psi$ - the amplitude of the wave itself -  and Max Born was the prime to achieve it. As we know, $\psi$ is solution of the famous wave equation for one particle with mass $m$, due to Schrödinger, 
\begin{equation}
 - \frac{{\hbar ^2 }}{{2m}}\nabla ^2 \psi  + V\psi  = i\hbar \frac{{\partial \psi }}{{\partial t}},
\end{equation}
where $V$ is the potential and $\hbar$ is the Planck constant divided by $2\pi$. The functions $\psi$ are in general complex. The connection of such quantities to the "real" world (or, which came to be the same, the acquirement of quantities called "observables") is represented by means of operations such as $\psi _i^* \psi _i$, where $\psi _i^*$ is the complex conjugate of $\psi _i$.
It prays a fundamental postulate of quantum mechanics that $\left| {\psi \left( {r,t} \right)} \right|^{\rm{2}} {\rm{ =  }}\psi {\rm{.}}\psi {\rm{ }}$ is the density of the probability ${P\left( {r,t} \right)}$ for a particle of mass $m$ to be found at the point $r$, on time $t$. Therefore, the likelihood to locate the particle inside an infinitesimal volumn of space $d\tau$ on that time $t$ is, 
\begin{equation}
\left| {\psi \left( {r,t} \right)} \right|^{\rm{2}} d\tau.
\end{equation}
Only in the case of one particle, the configuration space of the function $\psi$ is isomorphic to the tridimensional space of positions. For two particles, for example, the wave function of the system, $\psi(r_1,r_2)$, is defined in a configuration space of six dimensions.
Since the summation of the probabilities referring to events that are mutually exclusive is 1, it follows,
\begin{equation}
\int {P\left( {r,t} \right)} d\tau  = \int {\left| {\psi \left( {r,t} \right)} \right|^{\rm{2}} d\tau  = 1}. 
\end{equation}
Once $\psi$ is not an observable quantity, there is a certain freedom of choice of its form. Besides,  the solutions of linear equations, like Schrödinger's equation, may be multiplied by complex numbers, remaining solutions, so that expression (3) turns possible to choice a correct amplitude factor.
The point of view of the physical interpretation sustains that the probability $P(r,t)$ is in fact the reflection of an objective property of the "particle", which is that the possible eigenvalues coexist as propentions in a reference class until a macroscopic intervention (a measurement) takes place. Such intervention changes drastically the original reference class. Let us take a system with states $\left| \psi  \right\rangle$ and $\left| {\psi '} \right\rangle$ respectively before and after the experimental intervention. It's clear that the function $\psi$ is somewhat conjectural here, but, for all theorectical purposes, is ever possible to think this function as a set of states reducible to an unique state (the reduction of the "wave packet"). We must consider the set $\left| \psi  \right\rangle$ while not especified any function $\psi$ by the apparatus of measurement, in such manner that we have two distinct instances of the reality, one before and other after the observation. 
Quantum measurements are represented by a collection $\lbrace$ {\o$_k$} $\rbrace$ of operators that act upon the phase space of the system under observation. The subindex $k$ labels the possible results of measurement. Let us suppose that the system is in the initial state $\psi$. The probability for a certain state $k$ after the measurement is,
\begin{equation}
P(k) = \left\langle \psi  \right| \text{\o}_k^\dag \text{\o}_k \left| \psi  \right\rangle, 
\end{equation}
where $\text{\o}_k^\dag$ is the transposed conjugated of $\text{\o}_k$, $\left\langle {\psi |\psi } \right\rangle = 1$ and $\sum\limits_k {\text{\o}_k^\dag \text{\o}_k}= 1$.
Let it be $\left| {\varepsilon _1 } \right\rangle ,\left| {\varepsilon _2 } \right\rangle ,\left| {\varepsilon _3 } \right\rangle ...\left| {\varepsilon _q } \right\rangle$ an orthonormal base. So,
\begin{equation}
\text{\o}_k = \left| {\varepsilon _k } \right\rangle \left\langle {\varepsilon _k } \right|
\end{equation}
is a quantum measurement. The intervention of the apparatus modifies the state of the system to,
\begin{equation}
\frac{{\text{\o}_k \left| \psi  \right\rangle }}{{\sqrt{{\left\langle \psi  \right| \text{\o}_k^\dag \text{\o}_k \left| \psi  \right\rangle }} }} = \left| {\psi '} \right\rangle; 
\end{equation}
\begin{equation}
\frac{{\left| {\varepsilon _k } \right\rangle \left\langle {\varepsilon _k } | \psi  \right\rangle }}{{\sqrt{{\left\langle \psi  \right| \text{\o}_k^\dag \text{\o}_k \left| \psi  \right\rangle }} }} = \left| {\psi '} \right\rangle; 
\end{equation}
\begin{equation}
\frac{{\left| {\varepsilon _k } \right\rangle \left\langle {\varepsilon _k } | \psi  \right\rangle }}{{\sqrt{{\left\langle \psi | \left\langle {\varepsilon _k |\varepsilon _k } \right\rangle | {\varepsilon _k } \right\rangle \left\langle {\varepsilon _k } | \psi  \right\rangle }} }} = \left| {\psi '} \right\rangle; 
\end{equation}
\begin{equation}
\frac{{\left| {\varepsilon _k } \right\rangle \left\langle {\varepsilon _k } | \psi  \right\rangle }}{{\sqrt{{\left\langle \psi | {\varepsilon _k } \right\rangle \left\langle {\varepsilon _k } | \psi  \right\rangle }} }} = \left| {\psi '} \right\rangle; 
\end{equation}
\begin{equation}
\frac{{\left| {\varepsilon _k } \right\rangle \left\langle {\varepsilon _k } | \psi  \right\rangle }}{{{ | \left\langle {\varepsilon _k } | \psi  \right\rangle |} }} = \left| {\psi '} \right\rangle.
\end{equation}
Finally, the implementation of symmetries in generalized quantum mechanical coordinates \footnote{A symmetry in quantum mechanics is a discrete transformation or a group of continuous transformations that let invariant  the Hamiltonian (or the Lagrangian) and the canonical commutation relations of the system.} may be represented by a unitary operator in the Hilbert space \footnote{The vector space of quantum mechanics is a Hilbert space, that is, an orthonormal vector space in which
\begin{enumerate}
\item the vector components are complex scalars;
\item the scalar product satisfies
$\left\langle {\psi |\psi } \right\rangle  > 0$ for  $\left| \psi  \right\rangle  \ne 0$ otherwise $\left\langle {\psi |\psi } \right\rangle  = 0$;
\item if $a$ and $b$ are complex scalars, them\\
$\left\langle \chi  \right.|\left. {a{\psi _1} + b{\psi _2}} \right\rangle  = a\left\langle {\chi |{\psi _1}} \right\rangle  + b\left\langle {\chi |{\psi _2}} \right\rangle$;
\item the space is complete in the norm $\left\| \psi  \right\| = \sqrt {\left\langle {\psi |\psi } \right\rangle }$.
\end{enumerate}} , so that, 
\begin{equation}
\mathcal{U} ,\mathcal{U} ^\dag  \mathcal{U}  = \mathds{1},\left[ {H,\mathcal{U} } \right] = 0;\nonumber
\end{equation}
for the groundstate of the Hamiltonian $H$,
\begin{equation}
\mathcal{U} \left| {\psi _0 } \right\rangle  = \left| {\psi _0 } \right\rangle {\rm{  (not \; obvious!)}};\nonumber
\end{equation}
\begin{equation}
H\left| {\psi _0 } \right\rangle  = E_0 \left| {\psi _0 } \right\rangle ; \nonumber 
\end{equation}
\begin{equation}
\mathcal{U} H\left| {\psi _0 } \right\rangle  = E_0 \left( {\mathcal{U} \left| {\psi _0 } \right\rangle } \right).\nonumber
\end{equation}
In fact, accordingly the von Neumann theorem, a coordinate transformation that corresponds to a symmetry of the Hamiltonian let invariant the canonical commutation relations of the system and (here is the power of the theorem) may always be implemented by an unitary manner in the Hilbert space of the states. So,
\begin{equation}
\hat q'_i  = \mathcal{U}\left( \mathtt{S} \right)\hat q_i \; \mathcal{U}^\dag  \left( \mathtt{S} \right) = \mathtt{S}_{ij} \hat q_j; \nonumber 
\end{equation}
\begin{equation}
\mathcal{U}\left( \mathtt{S} \right) = e^{i\omega \mathtt{\hat O}}, \nonumber 
\end{equation}
where $ \mathtt{\hat O}$ is an operator that defines a motion constant (thereby furnishing good quantum numbers for the states of the system) so that $\mathtt{\hat O}^\dag =\mathtt{\hat O}$, and $\omega$ is the set of parameters defining the matrix $\mathtt{S}$.
Of course, as an effect of the macroscopic intervention, $\varepsilon _k$ shows some classic traces inherited from the apparatus. But quantum mechanics says nothing about de world out of the experiment. In particular, with respect to gravity, an approach by quantum field theory would need 1) an understandable model of gravitation accordingly some quantization algorithm applied to general relativity, which seems little bearing, and 2) an experimental frame able to reproduce the physical conditions under which the hypothetical quantum nature of gravity may come about, such as in a black hole singularity. In fact, one reason to brush aside an experimental program in this way is the difficulty of formulating quantum theory in a cosmological context in which the observers must be part of the system. Although it appears out of the blue, we may suppose there is a real messenger of gravity and imagine a "metaframe" to render gravitation in a familiar figurative language with no {\it a priori} concerns whether the messenger and its supersymmetric partner follow Bose or Fermi statistics beneath lab apparatus. This is my proposal: a supersymmetric meta-field theory on gravity. So, I define meta-field theory as a theory that introduces a supersymmetric metaframe to describe fields as sets of particular transformations between two types of entities, the supersymmetric partners in focus.
   
It seems to be the best moment to discuss in short semantic features of the explaining representation. An accurate investigation in supergravity, as well in general field theory, is already hard enough to aggregate more difficulties caused by conceptual mistakes. I will try, so much as possible, to refine the presentation of the formalism, provided is a great bind in SUSY literature the time lost with the nearly always confused roll of notations and conventions.

\subsection*{Recollection on general field theory}
\addcontentsline{toc}{subsection}{Recollection on general field theory}
 The classical approach of a system with symmetry is well known in elementary field theory. So, let us consider three complex fields changing beneath a type $\mathbf{U}(\mathds{1})$ symmetry accordingly the relations,
\begin{enumerate}
\item $\varphi ' = e^{i\alpha } \varphi$;
\item $\sigma ' = e^{ - i\alpha } \sigma$;
\item $\chi ' = e^{i\alpha } \chi$.
\end{enumerate}
The dynamics of these fields is assumed to be controlled by the Lagrangian density
\begin{equation}
\mathcal{L} = i\varphi ^ *  \dot \varphi  + i\sigma ^ *  \dot \sigma  + i\chi ^ *  \dot \chi  + i\varphi ^ *  \dot \chi  + i\chi ^ *  \dot \varphi  - \nonumber
\end{equation}
\begin{equation}
 - \frac{1}{{2m_1 }}\vec \nabla \varphi ^ *   \cdot \vec \nabla \varphi  - \frac{1}{{2m_2 }}\vec \nabla \sigma ^ *   \cdot \vec \nabla \sigma  - \frac{1}{{2m_3 }}\vec \nabla \chi ^ *   \cdot \vec \nabla \chi  - \nonumber
\end{equation}
\begin{equation}
 - \frac{1}{\mu }\vec \nabla \varphi ^ *   \cdot \vec \nabla \chi  - \frac{1}{\mu }\vec \nabla \chi ^ *   \cdot \vec \nabla \varphi  + \lambda \left( {\varphi ^ *  \varphi } \right)^2  + f\left( {\sigma ^ *  \sigma } \right)^2  + \nonumber
\end{equation} 
\begin{equation}
 + g\left( {\chi ^ *  \chi } \right)^2  + h\left( {\varphi ^ *  \chi } \right)^2  + h\left( {\chi ^ *  \varphi } \right)^2.
\end{equation}
The variation of the Lagrangian is given by,
\begin{equation}
\delta \mathcal{L} = 0 = \frac{{\partial \mathcal{L}}}{{\partial \varphi }}\delta \varphi  + \frac{{\partial \mathcal{L}}}{{\partial \dot \varphi }}\frac{\partial }{{\partial t}}\delta \varphi  + \frac{{\partial \mathcal{L}}}{{\partial \vec \nabla \varphi }} \cdot \vec \nabla \delta \varphi . 
\end{equation}
From this, we define a density $\rho$ and a current $j$ (see the Noether theorem at Chapter 2),
\begin{equation}
\frac{\partial }{{\partial t}}\underbrace {\left( {\frac{{\partial \mathcal{L}}}{{\partial \dot \varphi }}\delta \varphi } \right)}_\rho  + \vec \nabla  \cdot \underbrace {\left( {\frac{{\partial \mathcal{L}}}{{\partial \vec \nabla \varphi }}\delta \varphi } \right)}_j = 0.
\end{equation}
The derivatives of de Lagrangian are
\begin{equation}
\frac{{\partial \mathcal{L}}}{{\partial \dot \varphi }} = i\varphi ^ *   + i\chi ^ * ; 
\end{equation}
\begin{equation}
\frac{{\partial \mathcal{L}}}{{\partial \dot \chi }} = i\chi ^ *   + i\varphi ^ * ; 
\end{equation}
\begin{equation}
\frac{{\partial \mathcal{L}}}{{\partial \dot \sigma }} = i\sigma ^ * ; 
\end{equation}
\begin{equation}
\frac{{\partial \mathcal{L}}}{{\partial \vec \nabla \varphi }} =  - \frac{1}{{2m_1 }}\vec \nabla \varphi ^ *   - \frac{1}{\mu }\vec \nabla \chi ^ * ; 
\end{equation}
\begin{equation}
\frac{{\partial \mathcal{L}}}{{\partial \vec \nabla \varphi ^ *  }} =  - \frac{1}{{2m_1 }}\vec \nabla \varphi  - \frac{1}{\mu }\vec \nabla \chi .   
\end{equation}
Thus, the total density, that is, the density calculated and summed for all three fields is,
\begin{equation}
\rho  = i\left( {\varphi ^ *   + \chi ^ *  } \right)i\alpha \varphi  + i\sigma ^ *  \left( { - i\alpha \sigma } \right) + i\left( {\chi ^ *   + \varphi ^ *  } \right)i\alpha \chi ;
\end{equation}
\begin{equation}
\rho  = \alpha \left( { - \varphi ^ *  \varphi  - \chi ^ *  \varphi  + \sigma ^ *  \sigma  - \chi ^ *  \chi  - \varphi ^ *  \chi } \right) .
\end{equation}
The charge in classical level is defined by the integral of $\rho$ for the space-volumn, 
\begin{equation}
\mathcal{Q} = \int {d^3 } \vec x\left[ {\alpha \left( { - \varphi ^ *  \varphi  - \chi ^ *  \varphi  + \sigma ^ *  \sigma  - \chi ^ *  \chi  - \varphi ^ *  \chi } \right)} \right] .
\end{equation}
 The transition to the formalism of quantized fields redefines the fields as operators in Hilbert space, so that the charge is now,
\begin{equation}
\mathcal{Q}_{op} = \int {d^3 } \vec x\left( { - \varphi _{op}^ *  \varphi _{op}  - \chi _{op}^ *  \varphi _{op}  + \sigma _{op}^ *  \sigma _{op}  - \chi _{op}^ *  \chi _{op}  - \varphi _{op}^ *  \chi _{op} } \right).
\end{equation}
$\mathcal{Q}_{op}$ must be the operator which the eigenvalues are the values of the individual charges of the fields. All the operators ($\varphi$, $\sigma$, $\chi$, $\mathcal{Q}$) must act on vectors of state constructed from the {\it vacuum} by the action of creation operators in terms of which the fields are expanded,
\begin{equation}
\varphi  = \sum\limits_n {a_n } \psi _n ;
\end{equation}
\begin{equation}
\sigma  = \sum\limits_n {b_n } \eta _n ;
\end{equation}
\begin{equation}
\chi  = \sum\limits_n {c_n } \xi _n . 
\end{equation}
This brief review gives a resume of the basic ideas in general field theory, and it is enough for my pourposes in present essay.
\subsection*{On the invariance of the action in the rigid $N=1$ supersymmetry}
I want to exemplify how do we implement a SUSY transformation that lets invariant the action in an essentially temporal and unidimensional conjectured model. Having the usual approaching inspired in Nieuwenhuizen's work, we generally start from a real bosonic field $\varphi$ and a real fermionic field $\chi$, being $\chi(t)$ the customary independent Grassmann variable. In this explanation, I consider $t$ as the unique dimension, since in my theory time has precedence, and only $N = 1$ susy generators.
Grassmanian quantities obey the rule,
\begin{equation}
\chi (t_1 )\chi (t_2 ) =  - \chi (t_2 )\chi (t_1 ), \nonumber
\end{equation}
from which,
\begin{equation}
\frac{d}{{dt}}\left( {\chi (t)\chi (t)} \right) = \dot \chi (t)\chi (t) + \chi (t)\dot \chi (t) = 0 \to \chi (t)\dot \chi (t) =  - \dot \chi (t)\chi (t); \nonumber
\end{equation}
\begin{equation}
\frac{{\partial ^2 }}{{\partial t_1 \partial t_2 }}\left( {\chi (t_1 )\chi (t_2 )} \right) = \dot \chi (t_1 )\dot \chi (t_2 ) =  - \dot \chi (t_2 )\dot \chi (t_1 ). \nonumber
\end{equation}

For all $t$ we have,
\begin{equation}
\left\{ {\chi (t),\chi (t)} \right\} = \left\{ {\chi (t),\dot \chi (t)} \right\} = \left\{ {\dot \chi (t),\dot \chi (t)} \right\} = 0. \nonumber
\end{equation}
Also, the Hermitian conjugation on any Grassmanian quantities $\chi_i$ and $\chi_j$ reads,
\begin{equation}
\left( {\chi _i \chi _j } \right)^\dag   = \chi _j^\dag  \chi _i^\dag. \nonumber 
\end{equation}
So now, let us take the action given by,
\begin{equation}
S = \int {\mathcal{L}(t)dt}, \nonumber 
\end{equation}
to the simple Lagrangian,
\begin{equation}
\mathcal{L} = \frac{1}{2}\dot \varphi ^2  + \frac{i}{2}\chi \dot \chi.  \nonumber
\end{equation}
The Klein-Gordon action is expressed by the first term, while the Dirac action by the second. We mix fermions and bosons by the SUSY transformations,
\[
\begin{array}{l}
 \delta \varphi  = i\epsilon \chi ; \\ 
 \delta \chi  =  - \dot \varphi \epsilon . \\ 
 \end{array}
\]
The variation of the Lagrangian gives,
\[
\delta \mathcal{L} = \dot \varphi \delta \dot \varphi  + \frac{i}{2}\left( {\delta \chi \dot \chi  + \chi \delta \dot \chi } \right);
\]
\[
 \delta \mathcal{L}= \dot \varphi \delta \dot \varphi  + \frac{i}{2}\left[ { - \dot \varphi \epsilon \dot \chi  - \chi \left( {\ddot \varphi \epsilon  + \dot \varphi \dot \epsilon } \right)} \right];
\]
\[
 \delta \mathcal{L}= \dot \varphi \frac{d}{{dt}}\left( {i\epsilon \chi } \right) - \frac{i}{2}\left( {\dot \varphi \epsilon } \right)\dot \chi  - \frac{i}{2}\chi \frac{d}{{dt}}\left( {\dot \varphi \epsilon } \right);
\]
\[
 \delta \mathcal{L}= \dot \varphi i\dot \epsilon \chi  + \dot \varphi i\epsilon \dot \chi  - \frac{i}{2}\dot \varphi \epsilon \dot \chi  - \frac{i}{2}\frac{d}{{dt}}\left( {\chi \dot \varphi \epsilon } \right) + \frac{i}{2}\dot \chi \dot \varphi \epsilon {\rm{          }}; \]\\
Applying $\dot \chi \epsilon  =  - \epsilon \dot \chi$, it follows,
\[
 \delta \mathcal{L}= \dot \varphi i\dot \epsilon \chi  - \dot \varphi i\dot \chi \epsilon  + \frac{i}{2}\dot \chi \dot \varphi \epsilon  - \frac{i}{2}\frac{d}{{dt}}\left( {\chi \dot \varphi \epsilon } \right) + \frac{i}{2}\dot \chi \dot \varphi \epsilon {\rm{    }};
\]
\[
 \delta \mathcal{L} = \dot \epsilon \left( {i\dot \varphi \chi } \right) - \frac{i}{2}\frac{d}{{dt}}\left( {\chi \dot \varphi \epsilon } \right).
\]
The integration of the action gives,
\[
\int {\delta Ldt = } \left. { - \frac{i}{2}\chi \dot \varphi \epsilon } \right|_{ - \infty }^\infty   + i\int {\dot \epsilon \dot \varphi \chi } dt.
\]
Neglecting the boundary part at $t=\pm \infty$ and for $\dot \epsilon=0$, the action is invariant under the proposed SUSY transformations.
\subsection*{The axiomatics for a supersymmetric gravity}
\addcontentsline{toc}{subsection}{The axiomatics for a supersymmetric gravity}
As once told Mario Bunge, "Mathematical forms say by themselves nothing about material reality, and this is just why they may be used (in combination with semantic 'rules') to say so much about the external world. The eventual objective content that can be poured into mathematical forms lies entirely in the factual (physical, biological,...) meaning attached  {\it ad hoc} to the symbols appearing in them, that is, in the semantic 'rules'"\hspace{0.8 mm} (Bunge, 1979). I think that any lecture or class on advanced physics must begin with a great emphasis in this observation in order to safeguard one to overestimate symbols and representations. The development of gravitorial theory is replete of abstractions evoked by the search of representation means for the supersymmetric phenomenal scenario of gravitation. One of the axioms of the theory transcripted bellow shows this fact.\\
\\
\textbf{Axiom}
\\
\\
\textbf{$\dag$ Subset of original primitive base}
\begin{itemize}
\item $E_{adS}$ - anti-de Sitter space;
\item $\underline E$ - affine space;
\item $\mathbb{C}$ - Clifford algebra. 
\end{itemize}
\textbf{$\dag$ Proposition}
\\
\\
Given $E_{adS}$ beneath symmetry $\mathbb{O}(3,2) = \lbrace \exists \Upsilon, \Upsilon \in \mathbb{C}_{3,2}|\Upsilon^{t}\Upsilon = \mathbb{U}_{2}\rbrace$, there exists an affine space $\underline E$ such that for some $\Upsilon^{t} \in \mathbb{C}_{3,2}$ there is an affinor $|\left.G \right\rangle \in \underline E$ upon which it holds the tautomorphism $T :\Upsilon^{t}|\left.G \right\rangle \longmapsto |\left.G \right\rangle$, or, $\Upsilon^{t} (g^{11},g^{21}) = (g^{11},g^{21})$.
$$
$$
We consider the tautomorphism $T :\Upsilon^{t}|\left.G \right\rangle \longmapsto |\left.G \right\rangle$. It introduces a particular application of invariance valid for $|\left.G \right\rangle$ - type objects said "gravitors", the column vectors of the complex space upon which act Clifford algebra matrices. This axiom, presuming all the others of the theory, brings great normative meaning for the manner we must deal with gravitors. We note that such entities were arbitrarily symbolized by the sign $G$ aided by '$|$' and '$\rangle$'. Such set of signs, being or not associated to sounds, merely serves to give literal form to the ideas. So, while symbol, $G$ do not exists else as a possible tool of thinking within a wide semantic framework. But, ultrapassing the mute symbology and arriving to physics, what are gravitors? They are math constructs that need to be expressed in association with particular sentences in order to gain physical significance. Even so, they are only tools of thinking; they have not concrete existence. However the human understanding makes efforts to build some pictures about, we must to pretend nothing more than a logic consistence between the sentences and the transempiric reality they try to explain. Playing the role of tools for the rational activity, those constructs pitch a yarn concerning the world of the exterior things. And the more the "tale"\hspace{0.4 mm} is good, the more is its likelihood; the more we dismissed the rulers and the clocks, the more we persuade ourselves that only the physicists bestow sense to physics itself. Gravitors and all entities of contemporary physics give us the fugacious impression of naturalness when we divagate on the unfathomable secrets of the Universe. 

In the ambit of this ontological discussion, I would like to point out the very well known inquiry about the existence of the graviton. Tony Rothman and Stephen Boughn compiled an outstanding approach on this subject in the article "Can gravitons be detected?"\hspace{0.8 mm} based on Freeman Dyson's questioning whether it would be conceivable any experiment to detect a graviton in our quadridimensional Universe. Albeit the incredible frailness of the gravitational interaction, there must be possible to perform a device of detection, a plain of search, otherwise graviton becomes a metaphysical object and we will be forced to decide if it is acceptable to treat as physical this metaphysical entity. The authors say in the introduction of the paper that "For both physical and philosophical reasons the matter turns out to be not entirely trivial, and both considerations require that the rules of the game be defined at the outset. We concede at once that there appear to be no fundamental laws disallowing the detection of a graviton, and so we take the approach of designing thought experiments that might be able to detect one"\hspace{0.8 mm} (Rothman \& Boughn, 2006). I recommend the paper to the reader, but on the face of what I said in last paragraph the only question that makes sense is "what part of the experience may be assigned to the presence of gravitons?"\hspace{0.8 mm} or "what part of the experience is just representable by the construct 'graviton'?". We see that much more ingeniousness is required than conventional empirism. 
$$
$$
\begin{center}
$\diamondsuit\diamondsuit\diamondsuit$
\end{center}

\newenvironment{caixa}[1]
{\begin{center}
\fbox{\rule{1ex}{8ex}\hspace{1ex}{#1}\hspace{54ex}
}}
{\end{center}}

\newenvironment{caixa2}[1]
{\begin{center}
\fbox{\rule{1ex}{8ex}\hspace{1ex}{#1}\hspace{54ex}
}}
{\end{center}}

\newenvironment{caixa3}[1]
{\begin{center}
\fbox{\rule{1ex}{8ex}\hspace{1ex}{#1}\hspace{54ex}
}}
{\end{center}}

\newenvironment{caixa4}[1]
{\begin{center}
\fbox{\rule{1ex}{8ex}\hspace{1ex}{#1}\hspace{54ex}
}}
{\end{center}}

\newenvironment{caixa5}[1]
{\begin{center}
\fbox{\rule{1ex}{8ex}\hspace{1ex}{#1}\hspace{54ex}
}}
{\end{center}}

\newenvironment{caixa6}[1]
{\begin{center}
\fbox{\rule{1ex}{8ex}\hspace{1ex}{#1}\hspace{54ex}
}}
{\end{center}}

\newpage

\thispagestyle{empty}
\mbox{} 
\vfill
\begin{center}
\begin{flushleft}
\begin{caixa}
{\LARGE Chapter 1}
\end{caixa}
\end{flushleft}
\end{center}
\vfill
\vfill
\pagebreak

\section*{Representation: SUSY change partners in Wick-rotation}
$$
$$
\hrule
\markright{\bfseries Representation} 
\vspace{4ex}
Recalling basic principles, in general relativity a vector is an oriented object perfectly superposed at a particular point in spacetime, in such manner that to compare two vectors at distinct points it is necessary
to carry one over to the other in a certain way. Of great physical interest is the transport of a vector along a path with no turning or stretching, the so-called "parallel transport". This concept was introduced still in the early XX century by the Italian mathematician Tulio Levi-Civita, as an outcome of the growing need of formal simplifications. An important feature of this transport in curved spacetimes is that it depends on the path taken. In other words, a curved spacetime is one that, by definition, has $n$ different parallel transports between two points according to the $n$ possible paths. It means that if a point-particle, moving in the curved spacetime, is affected only by gravity, its velocity vector is parallel transported along the peculiar line it traces in the curved spacetime, that is to say, along the "geodesic".
\section*{Metric}
\addcontentsline{toc}{section}{Metric}
The representative spacetime in general relativity is a 4-dimensional Riemannian manifold (in fact, a pseudo-Riemannian manifold). It was Riemann who established the generic concept of multiplicity ({\it Mannigfaltigkeit}), later "manifold" in English version. A generic manifold $\mathpzc{M}^m$ is a topological space which is locally Cartesian, such that the calculus on it presupposes the existence of a smooth coordinate system at some neighborhood of each point. A Riemannian manifold is the pair $\mathpzc{R}^m:(\mathpzc{M}^m, g)$ where $g$ is a symmetric positive-definite bilinear form, a tensor field on $\mathpzc{M}^m$ called "metric" or "distance function", which satisfies the following axioms:
\begin{itemize}
\item for all $x_a ,x_b  \in \mathpzc{R}, g(x_a ,x_b ) \ge 0$;
\item if $g(x_a ,x_b ) = 0$, então $x_b = x_a$;
\item for all $x_a ,x_b  \in \mathpzc{R}, g(x_a ,x_b ) = g(x_b ,x_a )$. 
\end{itemize}
\section*{Weak gravitational waves}
Now, let us begin with gravitational radiation. Based on the theory of electromagnetic fields, it was proposed that, far from the source, gravitational waves must be described by a solution of a linear approximation of Einstein's field equations, so that the metric would be,
\begin{equation}
g_{\mu \nu }  = \eta _{\mu \nu }  + h_{\mu \nu }, 
\end{equation}
where $\eta _{\mu \nu }$ is a flat spacetime metric, and $h_{\mu \nu }$ is a correction such that $h_{\mu \nu }\ll1$. Applying harmonic coordinates, for $h=h_{\mu \nu }\eta ^{\mu \nu }=0$, we get, 
\begin{equation}
\frac{{\partial h_{\mu \nu } }}{{\partial x_\nu  }} = 0,
\end{equation}
and so Einstein's field equations for a vacuum take the form of the wave equation, that is,
\begin{equation}
\left( {\frac{{\partial ^2 }}{{\partial x_1^2 }} - \frac{{\partial ^2 }}{{\partial x_2^2 }} - \frac{{\partial ^2 }}{{\partial x_3^2 }} - \frac{{\partial ^2 }}{{\partial x_4^2 }}} \right)h_{\mu \nu }  = 0.
\end{equation}
Known the polarizations of the transversal wave, we have,
\begin{equation}
P_1  = \frac{1}{2}(h_{22}  - h_{33} ) 
\end{equation}
and
\begin{equation}
P_2  = h_{23}.
\end{equation}

Being $h_{23}= h_{32}$, we establish the notation,
\begin{equation}
h_{ +  - }  = h_{22}  - h_{33}
\end{equation} 
and
\begin{equation}
h_{ +  + }  = h_{23}  + h_{32}. 
\end{equation}

In resume, for a non-massive theory, we assume that a weak perturbation $h_{\mu \nu }$ travels across a flat background $\eta _{\mu \nu }$, far from field sources, as a plane wave in such manner that the metric is
$g_{\mu \nu }  = \eta _{\mu \nu }  + h_{\mu \nu }$. If we introduce a massive graviton, a direct consequence is the speed of propagation depending on the frequency, 
\begin{equation}
v(\omega ) = c\sqrt {1 - \frac{{m^2 c^2 }}{{\omega ^2 }}},
\end{equation}
and we fall into a bimetric theory to determine the six polarization modes of the gravitational wave for a massive graviton. The study of such modes is beyond the scope of this work; for all purposes, with the above speed expression in mind, I'll focalize mainly gravitor's formal structure associated to the wave propagation. 
\section*{Gravitons and gravitinos}
\addcontentsline{toc}{section}{Gravitons and gravitinos}
Gravitons, the hypothetical gauge bosons, messenger particles of gravity, are thought here as being associated to affinors deduced of an adS (anti-de Sitter) specific application. In the same way are thought not less hypothetical gravitinos, their supersymmetric fermionic gauge partners, commonly represented by Majorana vector spinor fields of spin 3/2. Gravitinos are expected to be present in all local supersymmetric models, which are regarded as the more natural extensions of the standard model of high energy physics. In the framework of minimal supergravity, the gravitino mass is, by construction, expected to lie around the electroweak scale, that is, in the $100 GeV$ range.

Supersymmetry describes fermions and bosons in a unified way as partners of a supermultiplet. Such multiplets necessarily have a decomposition in terms of boson and fermion states of different spins. So, the supergravity multiplet consists of the graviton and its superpartner, the gravitino (in fact, the gravitino multiplet contains (1; 3/2) and (-3/2; -1), that is a gravitino and a gauge boson; on the other hand, the graviton multiplet includes (3/2; 2) and (-2; -3/2), corresponding to the graviton and the gravitino). Really the graviton spin 2 derives from the rank 2 of the metric tensor $g_{\mu \nu }$ which describes the gravitational field. At first look, gravitino could have spin 5/2 as often as 3/2, but the advantage to choose spin 3/2 is the absence of the goldstino in supersymmetry breaking theories.

After formal considerations (Serpa, 2002), resulting components for gravitons and gravitinos, accordingly gravitorial representation, are respectively:
\vspace{8 mm}
\begin{equation}
 G_\mu = \left\{ {\left[ \begin{array}{l}
 \mathds{1}_2  \\ 
 \sigma _1  \\ 
 \end{array} \right],\left[ \begin{array}{l}
 \; \mathds{1}_2  \\ 
 i\sigma _2  \\ 
 \end{array} \right],\left[ \begin{array}{l}
 \; \mathds{1}_2  \\ 
 i\sigma _3  \\ 
 \end{array} \right],\left[ \begin{array}{l}
 \mathds{1}_2  \\ 
  \imath \! \mathbf{i}_2  \\ 
 \end{array} \right]} \right\},
\end{equation}
\begin{equation}
 g_\mu = \left\{ {\left[ \begin{array}{l}
 \; \imath \! \mathbf{i}_2  \\ 
 i\sigma _1  \\ 
 \end{array} \right],\left[ \begin{array}{l}
  \; \; \imath \! \mathbf{i}_2  \\ 
 {\scriptscriptstyle{-}}\sigma _2  \\ 
 \end{array} \right],\left[ \begin{array}{l}
  \; \; \imath \! \mathbf{i}_2  \\ 
 {\scriptscriptstyle{-}}\sigma _3  \\ 
 \end{array} \right],\left[ \begin{array}{l}
 \; \; \imath \! \mathbf{i}_2  \\ 
 {\scriptscriptstyle{-}}\mathds{1}_2  \\ 
 \end{array} \right]} \right\},
\end{equation}
where
\begin{equation}
 \mathds{1}_2 = \left( {\begin{array}{*{20}c}
   1 & 0  \\
   0 & 1  \\
\end{array}} \right),
\end{equation}
and
\begin{equation}
\imath \! \mathbf{i}_2 = \left( {\begin{array}{*{20}c}
   i & 0  \\
   0 & i  \\
\end{array}} \right),
\end{equation}
with the customary Pauli matrices,
\begin{equation}
\sigma _1  = \left( {\begin{array}{*{20}c}
   0 & 1  \\
   1 & 0  \\
\end{array}} \right),{\rm{ }}\sigma _2  = \left( {\begin{array}{*{20}c}
   0 & { - i}  \\
   i & 0  \\
\end{array}} \right),{\rm{ }}\sigma _3  = \left( {\begin{array}{*{20}c}
   1 & 0  \\
   0 & { - 1}  \\
\end{array}} \right). 
\end{equation}

We suppose the states of graviton are "mirrored" in states of gravitino, always in pairs, beneath adS Clifford algebra (gravitinos correspond to transformations of graviton affinors), so that,
\begin{equation}
g = \Upsilon G
\end{equation}
ou
\begin{equation}
G = \Upsilon ^ -  g \equiv  - \Upsilon g,
\end{equation}
being $\Upsilon$ and its inverse $\Upsilon ^ -$ elements of the algebra. These transformations are of type $O(n)$. We suppose too, remembering that nothing compel us to use anticommuting quantities to formulate symmetry between bosons and fermions, a superspace with four gravitorial coordinates in addition to the ordinary $x-y-z-t$. Also, as schematized in figure 2, there is a translation in spacetime when a particle (graviton) transforms into
its superpartner (gravitino) and then back into the original particle (graviton) (this is a fundamental property of supersymmetry). Such translation probably occurs in a jump between different metrics, which would be a reason of the difficulty to prove the existence of supergravity.
The content of equations (39) and (40) is in fact similar to the content of the supersymmetric transformations generated by quantum operators $\mathds{Q}$ that transmute fermionic states in bosonic ones and {\it vice-versa}:
\begin{equation}
\begin{array}{l}
 \mathds{Q}\left| {fermion} \right\rangle  = \left| {boson} \right\rangle ; \\ 
 \mathds{Q}\left| {boson} \right\rangle  = \left| {fermion} \right\rangle . \\ 
 \end{array}
\end{equation}
The column-vectors $\left| G \right\rangle$ above must obey a selective criterion derived from the theory in order to preserve solely the best logical representatives. This criterion is divided in three rules (the "iron trigon"):\\
1. Column vectors are linked to the adS Clifford Algebra,
\begin{center}
\begin{equation}
\left[ {\Upsilon ^a ,\Upsilon ^b } \right] = 2\eta ^{ab}\mathds{1},
\end{equation}
\begin{equation}
\left( {\Upsilon ^0 } \right)^2  = \mathds{1}, \Upsilon ^0  = \Upsilon ^{0 \dag}  {\rm{ }}({\rm{hermitian}}),
\end{equation}
\begin{equation}
\left( {\Upsilon ^a } \right)^2  = -\mathds{1}, (a = 1, 2, 3), \Upsilon ^a  = \Upsilon ^{a \dag} {\rm{ }}({\rm{a-}} {\rm{hermitian}}),
\end{equation}
\begin{equation}
\left( {\Upsilon ^4 } \right)^2  = \mathds{1}, \Upsilon ^4  = \Upsilon ^{4 \dag}  {\rm{ }}({\rm{hermitian}}),
\end{equation}
\begin{equation}
\Upsilon ^a \Upsilon ^b = - \Upsilon ^b \Upsilon ^a, a\neq b,
\end{equation}
\end{center}
by the tautomorphism $T :\Upsilon^{t}|\left.G \right\rangle \longmapsto |\left.G \right\rangle$, from which
\begin{equation}
\Upsilon ^t \left[ \begin{array}{l}
 g_{11}  \\ 
 g_{21}  \\ 
 \end{array} \right] = \left[ \begin{array}{l}
 g_{11}  \\ 
 g_{21}  \\ 
 \end{array} \right];
\end{equation}
2. Pair-wise selected column-vectors are physically representative iff each vector is a Wick-rotation of the other; \\
3. Column-vectors are elements of the semi-Abelian group $S(\mathds{1}, g)$ (special group of gravitors), deduced from the 16-second matrices of the vectors,
\begin{figure}[h]
\begin{center}
\includegraphics[scale=0.48]{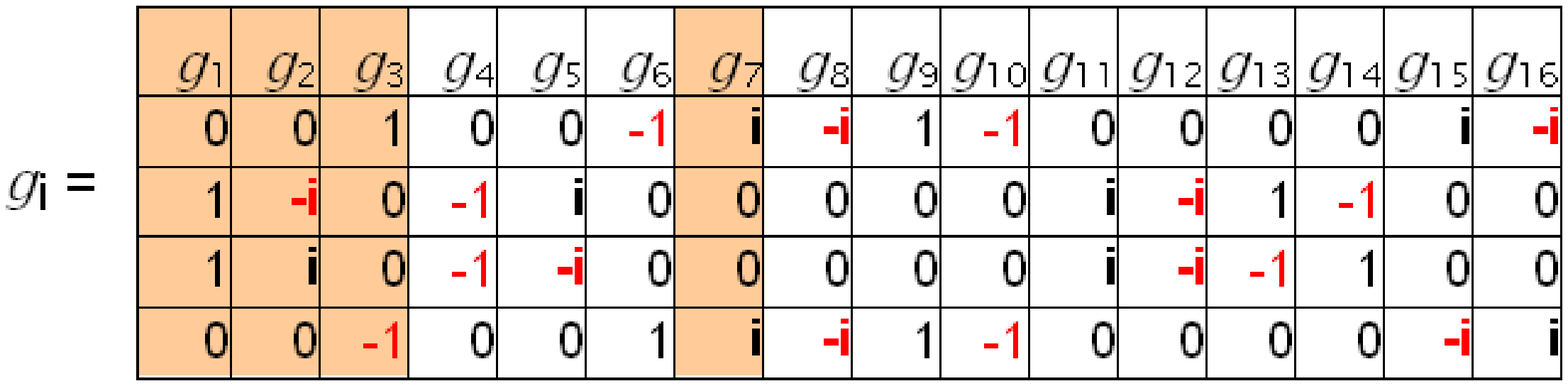} 
\end{center}
\end{figure}
where the four orange matrices integrate the column-generators $(\mathds{1}, g)$ of the group. The vectors $g_i$, which indeed form the group $S(\mathds{1}, g)$, compose by themselves the group $S(g)$ with a "feedback" property: $S(g)=iS(g)$. A program in R-language was made to test group elements and their creation by the generators. For example, with the group operation "$\otimes$" on all ordered $g_i$,\\
\begin{equation}
g_1  \otimes g_2  \otimes g_3  \otimes g_4  \otimes g_5  \otimes g_6  \otimes g_7  \otimes g_8  \otimes g_9  \otimes g_{10}  \otimes g_{11}  \otimes g_{12}  \otimes g_{13}  \otimes g_{14}  \otimes g_{15}  \otimes g_{16}  = g_{10}. 
\end{equation} 
\\
There are several motivations for the use of adS Clifford algebra, but it is enough to stand out that if one chooses a $2^n X 2^n$ matrix representation for the generators $\gamma^a$ of a simple $\mathbb{C}_{k,l}$ acting on spinor or gravitor space $\mathscr{G}$, so $\mathbb{C}_{k,l}$ is also generated by the transposed matrixes $
{\rm{\gamma }}_t^a  = {\rm B}{\rm{\gamma }}^a {\rm B}^{ - 1} 
$ (${\rm B}$ is uniquely defined)\footnote{For $\left[ {\gamma ^a \gamma ^b } \right] = 2g^{ab}$ and $p^a$ being components of a vector $p \in V$, ($V$ is a $2^n$ dimensional vector space endowed of a scalar product $g$), Cartan's equation says, for a Dirac spinor $\psi$, $\sum\limits_{a = 1}^{2n} {p^a \gamma _a } \psi  = 0$.}. Mathematical symmetries like this are very advantageous to deal with physical ones and they are generally taken with natural good receptivity. Besides, 1) Clifford algebras usually furnish spinorial representations of rotation groups and 2) supergravity does not exist in de Sitter space (Pilch, Sohnius \& Nieuwenhuizen, 1985), two sufficient reasons by which one applies adS Clifford algebra for supergravity with gravitorial affinors. Not so facile, however, is to use and justify physically the Wick-rotation. Its application is in such manner confuse that in many works I was even unable to resolve whether the authors were discussing having in mind Lorentzian or Euclidean signature; indeed, I could not see any clear justification with physical significance to introduce imaginary rotations in those discussions. Excepting the few one can find on Wick-rotation applied to the momentum variable $k_0$ in Green's functions, any remarkable reference has been done, over all about the possible roles fulfilled by bosons and fermions in Wick-rotation (concerning peculiarly the fermions, this was noted also by Nieuwenhuizen and Waldron)\footnote{P. van Nieuwenhuizen, A. Waldron, "A Continuous Wick Rotation for Spinor Fields and Supersymmetry in Euclidian Space", hep-th/9611043, proceedings of the string conference held at Imperial College, London, 1996.}. Even in early quantum physics, the imaginary unit is hollow of physical significance. As an example, concerning Pauli matrixes, we see that physicists like to put them in one-to-one correspondence with orthogonal directions in Euclidean 3-space, expressing their orthogonality by the Grassmannian outer product $\sigma _1  \wedge \sigma _2  = \sigma _1 \sigma _2  =  - \sigma _2 \sigma _1$. Thereby, the product $\sigma _1  \wedge \sigma _2  \wedge \sigma _3  = \sigma _1 \sigma _2 \sigma _3  = i$ reflects the identity between $i$, as the pseudo-scalar unit for Euclidean 3-space, and a trivector created by the outer product of the orthogonal vectors $\sigma _1$, $\sigma _2$ and $\sigma _3$. Everything is accepted tacitly as a mere formal result. Also in twistor theory, no direct physical interpretation is generally assigned to the complex coordinates. By the way, recalling Clifford algebras, we see that a reflection related to a plane orthogonal to $\gamma ^a$ is given by $\psi  \Rightarrow \gamma ^a \psi$ (see tautomorphism at Axiom 1) in spinor space, and for time-like $\gamma ^a$ we must substitute $\gamma ^a$ by $i\gamma ^a$ to satisfy the imposition of identity for squared reflections, indeed a beautiful feature but much more connected to mathematical modeling than to physical requests (I am trying simply to show where we may be more emphatic about physics). Finally, discussing Higgs mechanism for gravity and considering a Lorentz violating spectrum in a model for non-massive gravitons, scientists are once more laconic about the imaginary frequency at very low momenta. 
Only in 1977 there appear an interesting work of the French physicist and philosopher of science Jean Émile Charon, {\it Theorie de la Relativité Complexe}, in which he proposes a complex quadridimensional riemannian structure to the physical space with a metric,
\begin{equation}
Z_{\alpha \beta }  = \tilde Z_{\alpha \beta }  + i\mathord{\buildrel{\lower3pt\hbox{$\scriptscriptstyle\smile$}} 
\over Z} _{\alpha \beta } ,Z_{\alpha \beta }  = Z_{\beta \alpha ,} 
\end{equation}
so that,
\begin{equation}
ds^2  = Z_{\alpha \beta } dy^\alpha  dy^\beta ,  
\end{equation}
with
\begin{equation}
y^\alpha   = \tilde y^\alpha   + i\mathord{\buildrel{\lower3pt\hbox{$\scriptscriptstyle\smile$}} 
\over y} ^\alpha .  
\end{equation} 
Charon argues, among other ideas, that only such a complex space turns possible to extend general relativity to quantum field domain and justify the four complex extra components (including time) as a way to assign physical quantities (for example, the action associated to the spin) to the mathematical point of spacetime \footnote{J. Charon, "{\it Theorie de la Relativité Complexe}", Albin Michel, Paris (1977).}. The theory did not gain the merited attention, but it is a fact that Charon gave physical significance to imaginary dimensions.  

John G. Taylor, in his good times, ascribed clear physical sense to imaginary quantities when he wrote the third chapter of "The New Physics", entitled "Faster Than Light". Telling us about Einstein's famous article of 1905, we may read at page 94 of the Spanish version: "...{\it si acelerar una partícula hasta la velocidad de la luz exige una cantidad infinita de energía, acelerarla por encima de este valor requerería una energía imaginaria. Una cantidad imaginaria está formada por el producto de un número real y la raíz cuadrada de menos uno. Aun cuando esta cantidad puede manejarse sin mayor problema como símbolo, en la realidad no resulta posible medirla}" (Taylor, 1974). Here, an imaginary physical quantity is one to which there is no sense to apply rules or clocks; is one to which observational operations are not defined. Even so, it is considerably present within the frame that explains the world. Book criticisms apart, few times I saw so much clearness in a simple communication of an idea not so simple. 

Not exactly in the same reasoning line, but in a certain way similar to mine, Nieuwenhuizen and Waldron propose "a continuous Wick-rotation for Dirac, Majorana and Weyl spinors from Minkowski spacetime to Euclidean space, which treats fermions on the same footing as bosons" (Nieuwenhuizen \& Waldron, 1996). They emphasize that the study do not focuses the Wick-rotation of the momentum variable $k_0$ but a Wick-rotation of the field theory itself. After some observations, they were leaded to suggest for a Dirac spinor the Wick-rotation,
\begin{equation}
\Psi (\tau ,\vec x) \to \mathtt{S}(\theta )\Psi _\theta  (\tau ,\vec x),
\end{equation}
\begin{equation}
\Psi _{}^\dag  (\tau ,\vec x) \to \Psi _\theta ^\dag  (\tau ,\vec x)\mathtt{S}(\theta ),
\end{equation}
in which $\Psi _{\theta  = \pi /2}  \equiv \Psi _{\rm{E}}$ is the Euclidian Dirac spinor and $\mathtt{S}(\theta)$ a diagonal matrix with entries $(e^{\gamma ^4 \gamma ^5 \pi /2} ,1,1,1)$ that acts only Wick-rotating time sector (the exponents   $\gamma ^4$ and $\gamma ^5$ are elements of the Euclidean Clifford algebra). Resembling argumentations are applied to Majorana and Weil spinors.
In analogous sense, as we may define a dual field $\phi _D$ and a dual function $F_D \left( {\phi _D } \right)$, so that $\phi _D  = F'\left( \phi  \right)$ and $F'_D \left( \phi  \right) =  - \phi$ constitute a Legendre transformation $F_D \left( {\phi _D } \right) = F\left( \phi  \right) - \phi \phi _D$ or, which came to be the same, a duality symmetry,
\begin{equation}
\left( \begin{array}{l}
 \phi _D  \\ 
 \phi  \\ 
 \end{array} \right) \to \left( {\begin{array}{*{20}c}
   0 & 1  \\
   { - 1} & 0  \\
\end{array}} \right)\left( \begin{array}{l}
 \phi _D  \\ 
 \phi  \\ 
 \end{array} \right),
\end{equation} 
we have in gravitorial theory a symmetry,
\begin{equation}
\left( \begin{array}{l}
 \imath \! \mathbf{i}_2^{}  \\ 
 \sigma _\eta   \\ 
 \end{array} \right) \to \left( {\begin{array}{*{20}c}
   {\gamma ^ -  _{11} } & {\gamma ^ -  _{12} }  \\
   {\gamma ^ -  _{21} } & {\gamma ^ -  _{22} }  \\
\end{array}} \right)\left( \begin{array}{l}
 \imath \! \mathbf{i}_2^{}  \\ 
 \sigma _\eta   \\ 
 \end{array} \right)
\end{equation} 
for gravitinos or, 
\begin{equation}
\left( \begin{array}{l}
 \mathds{1}_2^{}  \\ 
 \sigma _\mu   \\ 
 \end{array} \right) \to \left( {\begin{array}{*{20}c}
   {\gamma _{11} } & {\gamma _{12} }  \\
   {\gamma _{21} } & {\gamma _{22} }  \\
\end{array}} \right)\left( \begin{array}{l}
 \mathds{1}_2^{}  \\ 
 \sigma _\mu   \\ 
 \end{array} \right)
\end{equation}
for gravitons.
What I will defend here is that Wick-rotation, a particular kind of abstract representation that serves very well to explain isometric transformations with physical significance, puts face-to-face, in a simple manner, the state-object and its qualitative change. It preserves a great mathematical inheritance, as we deal with the so intuitive and powerful idea of rotation group but, at the same time, warrants that "something" is no more the "same thing" from the viewpoint of system's physics under isometric isomorphism. When we Wick-rotate a bosonic representation, we are bringing a fermionic one but in a unique affine frame because $S(g)=iS(g)$. Graviton and gravitino share that affine space in pairs, as the object and its image.
$$
$$
\begin{center}
$\diamondsuit\diamondsuit\diamondsuit$
\end{center}

\newpage

\thispagestyle{empty}
\mbox{} 
\vfill
\begin{center}
\begin{flushleft}
\begin{caixa2}
{\LARGE Chapter 2}
\end{caixa2}
\end{flushleft}
\end{center}
\vfill
\vfill
\pagebreak

\section*{Lagrangian density and non-locality: backward to relics and beyond }
\addcontentsline{toc}{section}{Lagrangian density and non-locality: backward to relics and beyond}
$$
$$
\hrule
\begin{flushright} 
\small {\it Discovery of an elementary spin 3/2 particle in the laboratory\\ would be 
a triumph for supergravity because the only consistent field
theory for interacting spin 3/2 fields is supergravity.}
\\
\vspace{3.2mm}
$\mathpzc{Peter \hspace{0.4mm} van \hspace{0.4mm} Nieuwenhuizen}$
\end{flushright}

\markright{\bfseries Lagrangian density} 
\vspace{4ex} 
Any cosmological theory is suported by the gravitation theory. Gravity is the only relevant force in the scale of galaxy clusters and beyond. The gravitation theory can be constructed in different ways and this is still a source of puzzles, mainly in discussions about quantization of gravity and unification of the forces. In fact, there are three main approaches to relativistic gravity theories:
\begin{itemize}
\item gravity is a property of spacetime itself, the geometry of curved spacetime;
\item gravity is a kind of matter within the spacetime (the relativistic field theory in flat spacetime);
\item gravity is the effect of the direct interaction between ponderable particles.
\end{itemize}
No matter the choice, it is important to look upon that up to now relativistic gravity has been tested experimentally only in weak field approximation. All the well known relativistic effects as the delay of light signals, the gravitational lensing, the gravitational frequency shift, the pericenter advance, the rotational effects and the gravitational radiation from binary systems may be obtained from any reasonable gravitational theory. Also is important to regard that the shapes of the observable systems at large scales or even at galaxy scale are results of the long time cumulative action of gravity, which points to the relevance of the time in gravity phenomenon. Viewed by this angle, gravity is more a cosmic measure of the matter evolution all along the time. 
 First of all, in gravity theories one must consider that an ordinary rotation of a macroscopic device can be enough to disturb ontological status of a particle by emission of gravitational waves. This fact serves to detach the relevance that may have one of the simplest examples of a symmetry transformation. Besides, we know that orthogonal groups $O(k,l)$ - $k$ space entries and $l$ time entries - are of great usefulness because they describe Lorentz symmetry in spacetimes having diagonal metric with $k$ eigenvalues $+1$ and $l$ eigenvalues $-1$ (or vice-versa); on the other hand, I am interested in $O(k,l)$ Wick-rotations to represent the global entail between supergravity partners, and so I am not limiting the theory to an extended local supersymmetry with an $O(n)$ real internal symmetry.

As I first note, Wick rotations were introduced more like an ingenious math trick to regularize quantum field theories than a formal representation of a physical fact or property. Integrations on meromorphic functions over spacetimes of Minkowski, where the action $S$ appears in path integrals as $e^{-iS}$, are frequently faced from Feynman diagrams, and show divergence. Making the action imaginary, thus, becoming the metric Euclidean, the analytic continuation of $e^{ - iS}$ gains a real negative exponent, forcing path integral to converge. So, the Euclidean (E) transformation may be seen as,
\[
e^{i\int {d^4 x\left( { - \frac{1}{2}\partial _\mu  \phi \partial ^\mu  \phi } \right)} }  = e^{\int {d^4 x_E \left( { - \frac{1}{2}\partial _\mu ^E \phi \partial _E^\mu  \phi } \right)} }  \equiv e^{ - S_E },
\]
where we understand that $\mu  = 1,...,4$, and assume the Wick rotation,
\[
\begin{array}{l}
 t \to  - it \equiv x_4;  \\ 
 d^4 x =  - id^4 x_E.  \\ 
 \end{array}
\]
In gravitorial theory, Wick rotation acts to modify the physical status of the field; it does not enter the theory to remove mathematical divergences. Rather than viewing it like a trick for the convergence of the path integral, Wick rotation is treated as a necessary feature of the supersymmetric physics of gravity in the form of an enantiomorphic transformation.
$$
$$
\begin{center}
$\diamondsuit\diamondsuit\diamondsuit$
\end{center}
\begin{figure}
\begin{wrapfigure}[36]{l}[1mm]{4cm}
\includegraphics[scale=0.64]{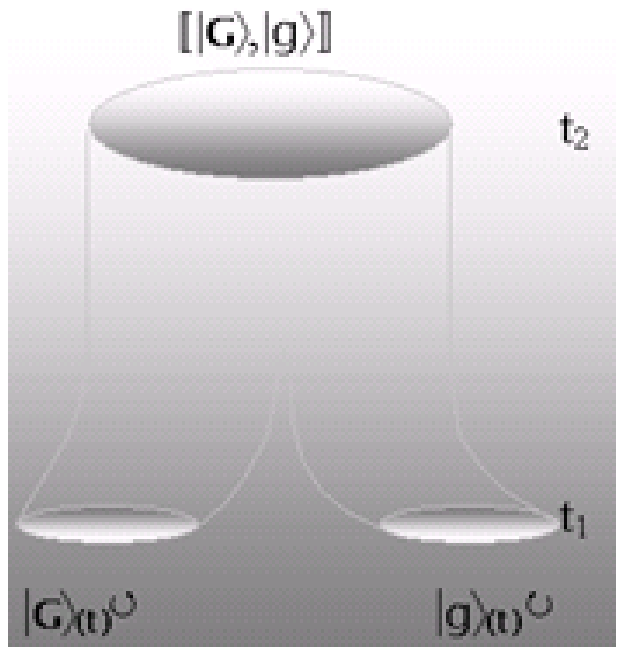} 
\small {{\it Fig.1- Scheme of two strings fusion as a cobordism. According to the theory, two chords, one fermionic and other bosonic, fuse with one another in order to make a loop of interaction. However I remember being so magnetized by string theories in the early 1990's, it is not the case to be here exhaustive about this subject. I'll just handle some introductory features related to the main thesis in view, remitting the reader to the last chapter of this work.}}
\end{wrapfigure}
The generators of SUSY are elements of the Clifford Algebra $\mathbb{C}_{3,2}$ and, at the same time, elements of the orthogonal group $O(k,l)$ that represent Wick-rotations when acting on gravitors; by its turn, the subset of selected gravitors is formed from elements of the group $S(\mathds{1}, g)$ which have the property to obey tautomorphism condition $T :\Upsilon^{t}|\left.G \right\rangle \longmapsto |\left.G \right\rangle$ for some component $\gamma^a$ of the algebra $\mathbb{C}_{3,2}$, except trivial quadratic-type forms $(\gamma^a)^2=\mathds{1}$. Since the reasons for a Wick-rotation go far beyond the simple operational commodity of an artifice, I must clarify its introduction with logic consistence.
In the early studies about gravitors, the former Lagrangian density that controls the interactions of the fields was, assuming all needed extra-constants included into kets,
\begin{equation}
\mathcal{L}_1  =  - 4\left( {\partial _\tau  \left| G \right\rangle \int {\left| g \right\rangle d\tau  + \partial _\tau  \left| g \right\rangle \int {\left| g \right\rangle d\tau } } } \right) - \Upsilon \left| G \right\rangle ^2 - \nonumber  
\end{equation}
\begin{equation}
- 2\left| G \right\rangle \left| g \right\rangle  - \Upsilon ^ -  \left| g \right\rangle ^2,
\end{equation}
with typical monoms of field products. Further application of the Noether theorem to all degrees of   freedom and some theoretical considerations leaded, within a topology $\mathbb{C}_{3,2}\times G \rightarrow \mathbb{C}_{3,2}^- \times g$, to 
\begin{equation}
\left[\kern-0.15em\left[ {\left| G \right\rangle ,\left| g \right\rangle } 
 \right]\kern-0.15em\right] = \partial _\tau  \left| G \right\rangle \int {\left| g \right\rangle d\tau }  - \partial _\tau  \left| g \right\rangle \int {\left| G \right\rangle d\tau }  = 0
\end{equation}
and to
\begin{equation}
\Upsilon \left| G \right\rangle ^2  = \Upsilon ^ -  \left| g \right\rangle ^2; 
\end{equation}
\begin{equation}
\left| G \right\rangle \not \partial \left| G \right\rangle  = \left| g \right\rangle \not \partial ^ -  \left| g \right\rangle, 
\end{equation}
with
\begin{equation}
\hspace{22 mm} \not \partial  = \Upsilon _\eta  \partial _\tau ^\eta . 
\end{equation}
\end{figure}

\newpage

Besides that Lagrangian of interactions, I considered another to exhibit a "gauge" field endowed of mass coupled to gravitino, such as,
\begin{equation}
\mathcal{L}_2 = M^2 |g\rangle \langle \mathord{\buildrel{\lower3pt\hbox{$\scriptscriptstyle\smile$}} 
\over G} |\partial _\tau  \langle \mathord{\buildrel{\lower3pt\hbox{$\scriptscriptstyle\smile$}} 
\over G} |\int {|g\rangle d\tau }  + 1/3 M^2 \langle \mathord{\buildrel{\lower3pt\hbox{$\scriptscriptstyle\smile$}} 
\over G} |^3  + \mathord{\buildrel{\lower3pt\hbox{$\scriptscriptstyle\smile$}} 
\over r} \partial _\tau  \mathord{\buildrel{\lower3pt\hbox{$\scriptscriptstyle\smile$}} 
\over r}, 
\end{equation} 
from which
\begin{equation}
\mathcal{L}_{Total}  =  - 4\left( {\partial _\tau  \left| G \right\rangle \int {\left| g \right\rangle d\tau  + \partial _\tau  \left| g \right\rangle \int {\left| g \right\rangle d\tau } } } \right) - \Upsilon \left| G \right\rangle ^2 - \nonumber
\end{equation}
\begin{equation}
- 2\left| G \right\rangle \left| g \right\rangle  - \Upsilon ^ -  \left| g \right\rangle ^2 + M^2 |g\rangle \langle \mathord{\buildrel{\lower3pt\hbox{$\scriptscriptstyle\smile$}} 
\over G} |\partial _\tau  \langle \mathord{\buildrel{\lower3pt\hbox{$\scriptscriptstyle\smile$}} 
\over G} |\int {|g\rangle d\tau }  + 1/3 M^2 \langle \mathord{\buildrel{\lower3pt\hbox{$\scriptscriptstyle\smile$}} 
\over G} |^3  + \nonumber
\end{equation}
\begin{equation}
+ \mathord{\buildrel{\lower3pt\hbox{$\scriptscriptstyle\smile$}} 
\over r} \partial _\tau  \mathord{\buildrel{\lower3pt\hbox{$\scriptscriptstyle\smile$}} 
\over r}. 
\end{equation}
Recalling conventions, a) in some situations it is convenient to express gravitational wave amplitude by the corresponding energy density flow in $erg/cm^2s$ or other bulk (this is not obligatory since we know the metric of spacetime), and b) relevant physical quantities have units which are powers of mass (for $\hbar = c =1$, length is $mass^{-1}$ because $\hbar c/mass$ is a length, etc). The fields $\left| G \right\rangle$ and $\left| g \right\rangle$, as coordinates of the whole system, are related to gravitons and gravitinos respectively. The field $
\langle \mathord{\buildrel{\lower3pt\hbox{$\scriptscriptstyle\smile$}} \over G} |$ is the gravitorial gauge inscription of the mass contribution outer to the adS zone (fig. 2) with $M^2$ appearing due to $
\langle \mathord{\buildrel{\lower3pt\hbox{$\scriptscriptstyle\smile$}} \over G} |$ and its coupling to other fields. The field ${\mathord{\buildrel{\lower3pt\hbox{$\scriptscriptstyle\smile$}} 
\over r} }$ is an auxiliary non-coupled field. Time integrals applied denote strong interference of system's history on local field inhomogeneities. Equations (59) and (60) show symmetries arising from stream evanescing superpositions as of the Noether current given by,
\begin{equation}
\frac{{\partial \mathcal{L}}}{{\partial \partial _4 |g\rangle }}\partial _4 |g\rangle  - \mathcal{L} = j.
\end{equation}
Never is overmuch to remind the content of Noether's theorem. For a system with Lagrangian density of the type $
\mathcal{L} = \mathcal{L}(\Phi ;\dot \Phi ,\vec \nabla \Phi )$, a continuous symmetry of $\mathcal{L}$ generates an equation of continuity $\frac{\partial }{{\partial \tau }}\rho  + \vec \nabla .\vec j = 0$, where $\rho$ and $\vec j
$ are functionals of $\Phi ,\dot \Phi ,\vec \nabla \Phi$, so that $Q = \int {d^3 \vec x\rho (\Phi ;\dot \Phi ,\vec \nabla \Phi )}$ is a constant of motion. 
Accordingly Noether's theorem, from the point of view of my theory, the fields and their derivatives or integrals are such that current ${\vec j}$ decais to $0$ faster than $1/\tau$ as $\tau \rightarrow \infty$. This is known as Noether's theorem for classic fields. We'll verify this property ahead (exercise 1). 
As pointed out by O'Raifeartaigh, "The Noether theorem gives the general relationship between symmetries and conservation laws. [...] Thus to every symmetry there corresponds a conserved quantity and conversely" (O'Raifeartaigh, 1997). This is clearly understood from (59) and (60), provided we focus time translations; Hamiltonian approach of system's dynamics also points out this evidence. 
 Let $\mathcal{P}$ being given by, 
\begin{equation}
\mathcal{P}=\frac{{\partial \mathcal{L}}}{{\partial \partial _4 |g\rangle }}.
\end{equation}
Hamiltonian density is defined as
\begin{equation}
\mathcal{H}=\mathcal{P}{\partial _4 |g\rangle } - \mathcal{L}.
\end{equation}
Now, $\mathcal{H}$ is precisely the content between brackets when we apply Noether's theorem in Lagrangians like (62). Let us take $\langle \mathord{\buildrel{\lower3pt\hbox{$\scriptscriptstyle\smile$}} 
\over G} | = \zeta e^{(\alpha _4 x_4  + ...)i}$. For $\partial _\tau  \langle \mathord{\buildrel{\lower3pt\hbox{$\scriptscriptstyle\smile$}}\over G} |$ we get:
\begin{equation}
\frac{d}{{d\tau }}\left\{ { - 1/3 M^2 \langle \mathord{\buildrel{\lower3pt\hbox{$\scriptscriptstyle\smile$}} 
\over G} |^3  - \mathord{\buildrel{\lower3pt\hbox{$\scriptscriptstyle\smile$}} 
\over r} \partial _\tau  \mathord{\buildrel{\lower3pt\hbox{$\scriptscriptstyle\smile$}} 
\over r} } \right\} = 0;
\end{equation}
\begin{equation}
\mathord{\buildrel{\lower3pt\hbox{$\scriptscriptstyle\smile$}} 
\over r} \partial _\tau  \mathord{\buildrel{\lower3pt\hbox{$\scriptscriptstyle\smile$}} 
\over r}  =  - iM^2 \alpha _4 \zeta ^3 \int {e^{3(\alpha _4 x_4  + ...)i} dt = }\nonumber 
\end{equation}
\begin{equation}
 =  - M^2 \zeta ^3 /3e^{3(\alpha _4 x_4  + ...)i}. 
\end{equation}
The same reasoning applied to the system of equations originated by $\mathcal{L}_1$ furnishes the symmetries (59) and  (60). The  conserved current there implies an inverse relation between $\mathbb{C}_{3,2} \times S(\mathds{1}, g)$ and its Wick-rotation. And more: a non-local algebraic structure $\left[\kern-0.15em\left[ {\left| G \right\rangle ,\left| g \right\rangle } \right]\kern-0.15em\right] =  0$.
Modern physicists do not like so much to deal with non-locality. There is no doubt it is a very strange attitude, since non-local phenomenology is very well-known in classical physics. In real space, incompressible turbulence is non-local as sound speed is infinite in incompressible fluids. In Fourier space, the equation also points to non-locality; on the other hand, the shell-to-shell energy transfer is local. The truth is a dialectical combination of both features. I can accept that locality is a logic necessity of the so-local human understanding. It makes possible the work of practical experimenters in order to obtain reproducible results. What I can not accept is the anthropocentric doctrine that build the world image by laboratory limitations. In present theory, supersymmetric transformations between gravitons and gravitinos occur beneath extreme and unstable conditions at the vicinity of a strong gravitational source. The vicinity is described by a very confined adS manifold where a tangent mass retention field ${\mathord{\buildrel{\lower3pt\hbox{$\scriptscriptstyle\smile$}} 
\over r} }$, the filtrino, acts to intermediate the mass exchange with Minkowsky space. The instability of adS world models (Coleman \& Luccia, 1980) is precisely the physical "spice" for a certain virtual mechanism of equilibrium with evasion of gravitons from the neighborhood of gravitational sources by adS - $O(k,l)$ Wick-rotations of gravitinos (remember the "iron trigon"). This is quite consistent with $1/r^3$ decay law of gravitino's influence, since adS zone tends to remain next to the source. As strong gravitational sources are in general very far, it would be a mistake to disregard historical roots in present ontological status of a graviton, whose lifetime is presumed very large but unknown. At the laboratory, the maximum one may suppose, if such is possible and after modest precautions to isolate experiments, is a situation in which similar initial conditions could be reproduced; thus, we gain,
\begin{equation}
\int {|g\rangle d\tau }  = 0,
\end{equation}
as well as
\begin{equation}
\int {|G\rangle d\tau }  = 0,
\end{equation}
and the reduced total Lagrangian,
\begin{equation}
\mathcal{L}_{Total}  =  - \Upsilon \left| G \right\rangle ^2 - 2\left| G \right\rangle \left| g \right\rangle - \nonumber
\end{equation}
\begin{equation}
- \Upsilon ^ -  \left| g \right\rangle ^2 + 1/3 M^2 \langle \mathord{\buildrel{\lower3pt\hbox{$\scriptscriptstyle\smile$}} 
\over G} |^3  + \nonumber
\end{equation}
\begin{equation}
+ \mathord{\buildrel{\lower3pt\hbox{$\scriptscriptstyle\smile$}} 
\over r} \partial _\tau  \mathord{\buildrel{\lower3pt\hbox{$\scriptscriptstyle\smile$}} 
\over r}. 
\end{equation}
But this sounds more as a {\it Gedankenexperiment} than a realistic one. The reality is, unfortunately, implacable. The weakness of gravitational force is extraordinary, and physicists are not able to directly verify the existence of gravitational waves, much less to simulate initial conditions for gravitorial transformations; gravitino interactions are of the rate of Newton's extremely small constant $G_N=0.67x10-38GeV^{-2}$. What makes time to be prevalent here is that gravity tends to manifest its effects only after long time of cumulative events; just because this feature we do not perceive gravitons or gravitational waves. The locality of human experience is inadequate to reproduce what occurred so long ago in extreme mass clustering conditions. Gravity is not a common force but a force that represents time strength in cosmic evolution at large scales. We need to admit the eventual impossibility of complete unification; we need to treat gravity contrarily to the usual misconception of another ordinary quantizable physical manifestation well localized in spacetime. The sole one can wait for is unification of rational thinking and, in a certain way, a {\it lato sensu} unification of applied general tools, perhaps the only kind we can expect to see beneath the logical objections pointed out by Ishan and colleagues. Transparence of gravitation with respect to all other forces seems to confirm this point of view.

In addition, it is more logic to imagine that gravitation operates by a timelike field evolution (non-local), so that we can actually measure only residual quantities associated to spacelike field evolution (local). To do this, we must find an appropriate "gauge", which gives a local response after incorporate the inheritance of the whole process (next section).
Now, consider an operator $\Im ={\rm{\gamma }}_\eta  \partial _\tau ^{ \eta } {\rm{ + \gamma }}_\eta  M$ acting on $\left| G \right\rangle ^2$ and $\left| g \right\rangle ^2$ such that
\begin{equation}
\Im \left| G \right\rangle ^2= 2\left| G \right\rangle \langle{\rm{\gamma }}_\eta  \partial _\tau ^{ \eta } \left| G \right\rangle {\rm{ + \gamma }}_\eta M_2 \left| G \right\rangle\rangle
\end{equation}
and
\begin{equation}
\Im ^- \left| g \right\rangle ^2= 2\left| g \right\rangle \langle{\rm{\gamma }}_\eta  ^ -  \partial _\tau ^{ \eta } \left| g \right\rangle {\rm{\gamma }}_\eta  ^ -  
 M_{3/2} \left| g \right\rangle\rangle.
\end{equation}
From (60) we know that holds the relation,
\begin{equation}
2\left| G \right\rangle \langle{\rm{\gamma }}_\eta  \partial _\tau ^{ \eta } \left| G \right\rangle \rangle= 2\left| g \right\rangle \langle{\rm{\gamma }}_\eta  ^ -  \partial _\tau ^{ \eta } \left| g \right\rangle \rangle .
\end{equation}
To connect both actions based upon figure 2, we must remind that gravitino local loss of mass under filtrino field rebates the mass attached to graviton, from which,
\begin{equation}
\left| g \right\rangle \langle{\rm{\gamma }}_\eta  ^ -  \partial _\tau ^{ \eta } \left| g \right\rangle \rangle - \langle {\rm{\gamma }}_\eta  ^ -  
 M_{3/2} \left| g \right\rangle^2 \rangle  = \nonumber
\end{equation}
\begin{equation}
\left| G \right\rangle \langle{\rm{\gamma }}_\eta \partial _\tau ^{ \eta } \left| G \right\rangle \rangle - \langle {\rm{\gamma }}_\eta (M_2 - {\mathord{\buildrel{\lower3pt\hbox{$\scriptscriptstyle\smile$}} 
\over \varepsilon } }) \left| G \right\rangle^2 \rangle ,
\end{equation}
where ${\mathord{\buildrel{\lower3pt\hbox{$\scriptscriptstyle\smile$}} 
\over \varepsilon } }$ is the retaining mass due to field, ${\mathord{\buildrel{\lower3pt\hbox{$\scriptscriptstyle\smile$}} 
\over r} }$. Thus,
\begin{equation}
{\rm{\gamma }}_\eta ^ - \left| g \right\rangle ^2  {\rm{ = \gamma }}_\eta  \frac{{ - (M_2  - \mathord{\buildrel{\lower3pt\hbox{$\scriptscriptstyle\smile$}} 
\over \varepsilon } )}}{{M_{3/2} }} \left| G \right\rangle ^2.
\end{equation}
Operator $\Im$ acts to ad a kinematical relation between graviton and gravitino if we consider massive gravitons in Wick-rotation at adS zone.
$$
$$
\begin{center}
$\diamondsuit\diamondsuit\diamondsuit$
\end{center}

\subsection*{From "non-local gauge" to "local coupling gauge"}
\addcontentsline{toc}{subsection}{From "non-local gauge" to "local coupling gauge"}
$$
$$
\hrule
\begin{flushright} 
\small {\it The final goal of supersymmetric meta-field theory is to provide\\ a consistent representation of gravity and at the same time to explain gravitation without any interference of the observer on the observed facts.}
\\
\vspace{3.2mm}
$\mathpzc{Nilo \hspace{0.4mm} Serpa}$
\end{flushright}
$$
$$
The introduction of a non-local inheritance factor as a device to include far-off interferences that may exist in certain phenomena is due to Vito Volterra, at the early twenty century, with his famous equation to the growth problem,
\begin{equation}
\frac{1}{y}\frac{{dy}}{{dt}} = a + by + \int_c^t {K(t,s)y(s)ds}. 
\end{equation}
Not much has been made to make good use of this very logical idea in the so-called "hereditary mechanics", a name coined by Emile Picard (to him, modern generations owe a famous work in three volumes, the Traité d'Analyse, 1891-96). Accordingly this great French mathematician, in all the study of classical mechanics "the laws which express our ideas on motion have been condensed into differential equations, that is to say, relations between variables and their derivatives. 
\begin{figure}
\begin{center}
\includegraphics[scale=0.56]{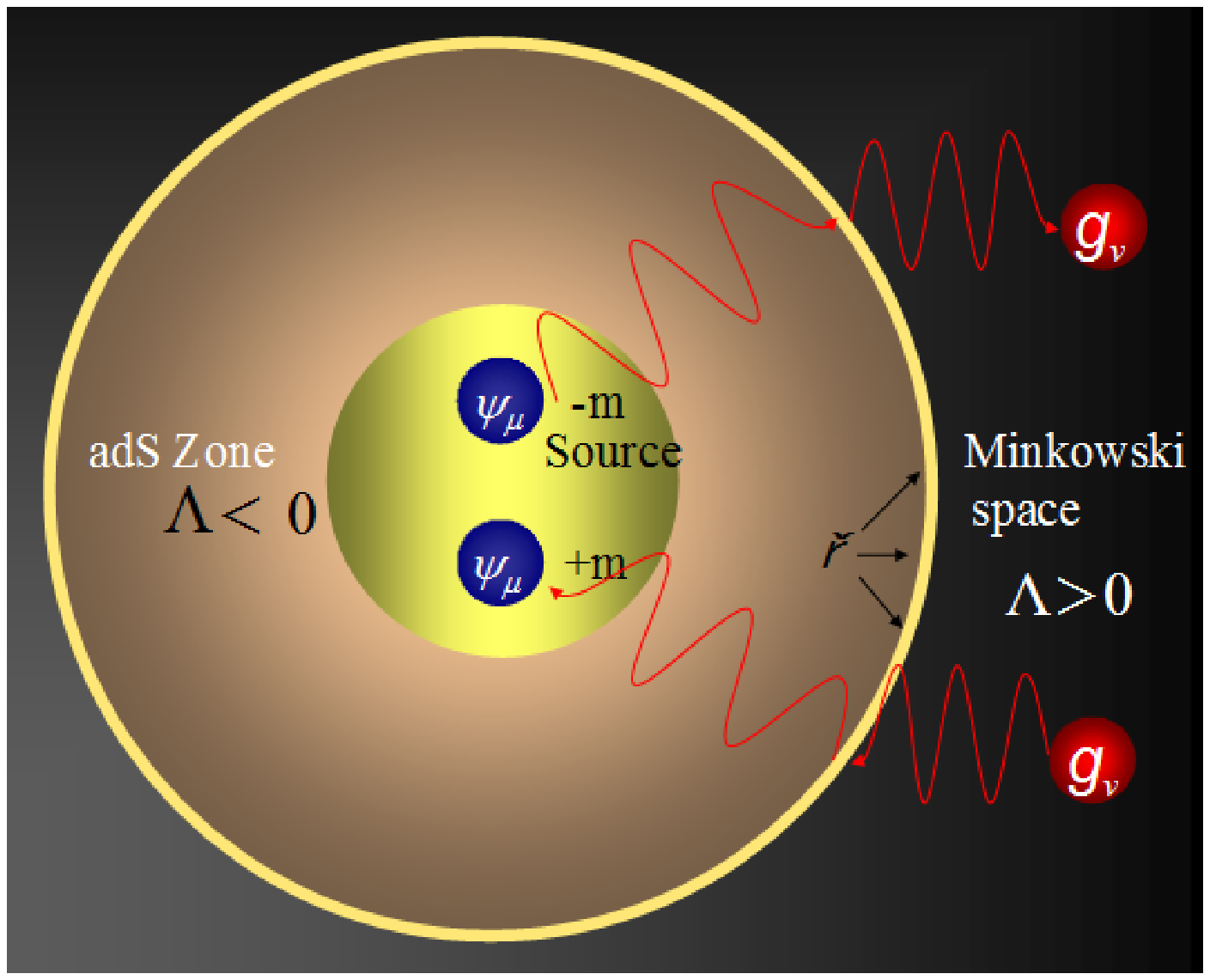}

\begin{quote}
\small{{\it Fig. 2 - Diagram of adS zone boundary interchange. Gravitinos $g$ lose mass smoothly to pass to Friedmann-Lemaître-Robertson-Walker (FLRW) space as gravitons $G$. On the inverse way, gravitons $G$ gain mass to cross  back to gravitational source vicinity as gravitinos $g$. From the viewpoint of a quantum field theory, with a vanishing cosmological constant we have a supersymmetry breakdown scale as a vacuum expectation value of auxiliary field $\mathord{\buildrel{\lower3pt\hbox{$\scriptscriptstyle\smile$}}\over r}$ just placed at the boundary. The width of adS zone depends on source intensity. The field $\mathord{\buildrel{\lower3pt\hbox{$\scriptscriptstyle\smile$}}\over r}$ plays a  similar part of the Higgs mechanism. I adopted a spherical Lemaître-Tolman (LT) interface between adS and FLRW zones for further considerations. The fundamental idea is to apply the Swiss cheese inhomogeneous model to the Universe until some scale limit to be determined. Several LT bubbles may contain a central massive adS zone that determines the intensity and diameter of the filtrino field boundary. The final hypothesis is that the gravitino survives mainly within the limits of LT zones, that is, it is most likely to find gravitinos inside LT zones.}}
\end{quote}
\end{center}
\end{figure}
We must not forget that we have, in fact, formulated a principle of nonheredity, when we suppose that the future of a system depends at a given moment only on its actual state, or in a more general manner, if we regard the forces as depending also on velocities, that the future depends on the actual state and the infinitely neighboring state which precedes. This is a restrictive hypothesis and one that, in appearance at least, is contradicted by the facts. Examples are numerous where the future of a system seems to depend upon former states. Here we have heredity. In some complex cases one sees that it is necessary, perhaps, to abandon differential equations and consider functional equations in which there appear integrals taken from a distant time to the present, integrals which will be, in fact, this hereditary part" (Picard, 1907). An absolute gravitation local theory seems to me so absurd as the old idea of absolute space. There is no sense regarding the sky without its evolutional non-local essence. Although the disregard of inheritance factors is in part consequence of an exaggeration of simplification, non-locality phobia in quantum field theory is very related with the fear to lose Lorentz and gauge invariance, both well preserved with local variables. We must remember that cubic string field theory (analogous to Chern-Simons gauge theory) \footnote{Chern-Simons gauge theory introduces a gauge field $A_\nu ^\mu  (x)$ and the corresponding Lagrangian $\mathcal{L} = \frac{k}{{4\pi }}\int {Tr(A \wedge {\rm{d}}A + \frac{2}{3}A \wedge A \wedge A)}$. As no metric is necessary, it is generally covariant.} given by,
\begin{equation}
S = \int {A*QA{\rm{ + }}\frac{2}{3}A*A*A}, 
\end{equation}
is invariant under the gauge transformation,
\begin{equation}
A \to A + Q\mathit{\Lambda} + \left[ {A, \mathit{\Lambda} } \right],
\end{equation}
with $\delta A = Q\mathit{\Lambda} + \left[ {A, \mathit{\Lambda} } \right]$.
\subsection*{Without jeopardizing clarity: diffeomorphism instead of coupling gauge}
\addcontentsline{toc}{subsection}{Without jeopardizing clarity: diffeomorphism instead of coupling gauge}
One may note that equation (59) tells that differences of local field transformations under field relics of the respective partners (interaction loops) are independent of the spacetime section (a flavor of Lorentz invariance). It is supposed that gauge invariant operators of non-local field theories are hard to write down. Perhaps one may believe in this 1) only thinking about an arbitrary gauge introduction with no much attention for the fact that to search a gauging of a global symmetry is to search a new interaction, and 2) forgetting the more general approach provided by diffeomorphic transformations. I derived a transformation from Euler equation applying to expression (62), the equation of gravitino's mass acquisition, according to which,
\begin{equation}
\frac{d}{{d\tau }}\left\{ {M^2 |g\rangle \langle \mathord{\buildrel{\lower3pt\hbox{$\scriptscriptstyle\smile$}} 
\over G} |\int {|g\rangle d\tau } } \right\} - M^2 |g\rangle \partial _\tau  \langle \mathord{\buildrel{\lower3pt\hbox{$\scriptscriptstyle\smile$}} 
\over G} |\int {|g\rangle d\tau }  -  M^2 \langle \mathord{\buildrel{\lower3pt\hbox{$\scriptscriptstyle\smile$}} 
\over G} |^2  = 0;
\end{equation}
\begin{equation}
\langle \mathord{\buildrel{\lower3pt\hbox{$\scriptscriptstyle\smile$}} 
\over G} | = |g\rangle ^2  + \partial _\tau  |g\rangle \int {|g\rangle d\tau }. 
\end{equation}
Expression (81) is said to be a "non-local gauge" (at first called "translocal gauge"). From a Lagrangian density with historical terms we obtain a suitable transformation of type $A \to A' = A^2 + \partial _\tau  A\int {Ad\tau } 
$, and we may prove that, 
\begin{equation}
A' = A^2  + \partial _\tau  A\int {Ad\tau }  = A^2  - \frac{1}{{\alpha ^2 }}(\partial _\mu   A)^2  = 
\end{equation}
\begin{equation}
=(A - \frac{1}{{\alpha ^{} }}\partial _\mu   A).(A + \frac{1}{{\alpha ^{} }}\partial _\mu   A).
\end{equation} 
In fact, this transformation maps an object into another whose  locality  is  arrested  from  far  away  in  time; we  are dealing more precisely with a diffeomorphism $D: A \rightarrow D(A)$, as the map $D$ is invertible and $C^{\infty}$.
After this singular "gauge product" of the sum by the difference, I start with a simple integro-differential equation,  
\begin{equation}
 - \frac{1}{{\alpha ^2 }}(\partial _\mu   A)^2  = \partial _\tau  A\int {Ad\tau }. 
\end{equation}
The left-hand side is the spacelike (local) remainder of field evolution, while the right-hand side is the instantaneous field status under influence of the field history (non-local). For the sake of brevity, taking one spatial dimension solely, a solution is,
\begin{equation}
A = {\rm A}e^{i(\alpha {\mu } + \beta {\tau })} \left( \begin{array}{l}
 \imath \! \mathbf{i}_2^{}  \\ 
 \sigma _\eta   \\ 
 \end{array} \right);
\end{equation}
\begin{equation}
A_\nu ^\kappa   = {\rm A}^\kappa  e^{i(\alpha {\mu } + \beta {\tau })} \sigma _\nu .  
\end{equation}
This solution is nothing more than the "shadow" gravitational wave associated to gravitino's polarization. Globally, we may Wick-rotate to graviton field,
\begin{equation}
A_\eta ^\kappa   = \gamma _\eta ^\nu  {\rm A}^\kappa  e^{i(\alpha {\mu } + \beta {\tau })} \sigma _\nu .
\end{equation}
In present model, on the contrary of Nieuwenhuizen's model, continuous Wick-rotation is a global consequence of the wave oscillation encapsulated by rotating affinors. An advantage is the automatic generalization to odd dimensions, since time is "flowing" in the action and Wick-rotation is not necessarily function of $\gamma^5$. Of course, extra-dimensions are not available here because the "iron trigon" and not due to any restriction about Wick-rotation itself.
To verify the invariance of the Lagrangian (the preservation of the physics) beneath transformation (81) we may rewrite, 
\begin{equation}
\mathcal{L} = {\rm{ }}M^2 |A\rangle \langle A^2 |\partial _\tau  \langle A^2 |\int {|A\rangle d\tau  + 1/3 M^2 \langle A^2 |^3  + \mathord{\buildrel{\lower3pt\hbox{$\scriptscriptstyle\smile$}} 
\over r} \partial _\tau  \mathord{\buildrel{\lower3pt\hbox{$\scriptscriptstyle\smile$}} 
\over r} }, 
\end{equation}
and, after (81),
\begin{equation}
\mathcal{L} = {\rm{ }}M^2 |A\rangle \langle A'|\partial _\tau  \langle A'|\int {|A\rangle d\tau  + 1/3 M^2 \langle A'|^3  + \mathord{\buildrel{\lower3pt\hbox{$\scriptscriptstyle\smile$}} 
\over r} \partial _\tau  \mathord{\buildrel{\lower3pt\hbox{$\scriptscriptstyle\smile$}} 
\over r} }= \nonumber
\end{equation}
\vspace{0.1in}
\begin{equation}
 = M^2 |A\rangle \langle (A - 1/\alpha \partial _\mu   A).(A + 1/\alpha \partial _\mu   A)|\partial _\tau  \langle (A - 1/\alpha \partial _\mu   A).(A + 1/\alpha \partial _\mu   A)|. \nonumber
\end{equation}
\vspace{0.16in}
\begin{equation}
.\int {|A\rangle d\tau  + } 1/3 M^2 \langle (A - 1/\alpha \partial _\mu   A).(A + 1/\alpha \partial _\mu   A)|^3  + \mathord{\buildrel{\lower3pt\hbox{$\scriptscriptstyle\smile$}} 
\over r} \partial _\tau  \mathord{\buildrel{\lower3pt\hbox{$\scriptscriptstyle\smile$}} 
\over r}. 
\end{equation}
Calculation shows invariance by,
\begin{equation}
\mathcal{L} = \frac{7}{3}M^2 A^6 e^{6i(\alpha {\mu } + \beta {\tau })} \left( \begin{array}{l}
 \imath \! \mathbf{i}_2^{}  \\ 
 \sigma _\eta   \\ 
 \end{array} \right) + \mathord{\buildrel{\lower3pt\hbox{$\scriptscriptstyle\smile$}} 
\over r} \partial _\tau  \mathord{\buildrel{\lower3pt\hbox{$\scriptscriptstyle\smile$}} 
\over r}, 
\end{equation}
\begin{equation}
\mathcal{L}' = \frac{{32}}{3}M^2 A^6 e^{6i(\alpha {\mu } + \beta {\tau })} \left( \begin{array}{l}
 \imath \! \mathbf{i}_2^{}  \\ 
 \sigma _\eta   \\ 
 \end{array} \right) + \mathord{\buildrel{\lower3pt\hbox{$\scriptscriptstyle\smile$}} 
\over r} \partial _\tau  \mathord{\buildrel{\lower3pt\hbox{$\scriptscriptstyle\smile$}} 
\over r}. 
\end{equation}
All told, a diffeomorphism, 
\begin{equation}
D:A \to D(A) = (A - \frac{1}{{\alpha ^{} }}\partial _\mu   A).(A + \frac{1}{{\alpha ^{} }}\partial _\mu   A)
\end{equation}
is analogous in certain respects to a gauge transformation, except by the intrinsically non-local statement,
\begin{equation}
- \frac{1}{{\alpha ^2 }}(\partial _\mu   A)^2  = \partial _\tau  A\int {Ad\tau }. \nonumber 
\end{equation}
This is an important ingredient for a realistic theory of supergravity and hereafter one must understand "diffeomorphic invariance" when reads "gauge invariance", unless indication in contrary. By this angle, we may say that supergravity is a diffeomorphic theory of global supersymmetry with gravitorial character. In other words, if supersymmetry is, so to speak, a gauge symmetry with spinorial character, accordingly meta-field theory supergravity is a diffeomorphic symmetry with gravitorial character.  
In most of the situations, conventional gauge-invariance is required only to the Lagrangian, but the action is physically more significant if we think the field system with dynamical concerns. Unfortunately, sometimes we deal with circumstances in which there is no complete gauge-invariance for both magnitudes. Let us take the example of the massive Yang-Mills Lagrangian for a well-grounded report of the massive non-Abelian gauge field dynamics. In use of independent dynamical variables $A_T^{\alpha \mu } (x)$,
\begin{equation}
\mathcal{L} =  - \frac{1}{4}F_T^{\alpha \mu \nu } F_{T\mu \nu }^\alpha   + \frac{1}{2}M^2 A_T^{\alpha \mu } A_{T\mu }^\alpha ,  
\end{equation} 
where the first term is a gauge-invariant used to describe the dynamics of the fields interacting with one another. The second term is not gauge-invariant and serves to determine kinematic features of the fields. In spite of this gauge-invariance of the massive term, the action referred to $\mathcal{L}$ is invariant with respect to gauge transformations of the type, 
\begin{equation}
\delta A_{T\mu }^a  = D_{T\mu }^{ab} \theta ^b , 
\end{equation} 
with
\begin{equation}
D_{T\mu }^{ab}  = \delta ^{ab} \partial _\mu   - gf^{abc} A_{T\mu }^c.
\end{equation}
In our case, both Lagrangian and action ensure gauge-invariance, a situation that may happen for massive Abelian gauge fields. This shows that, at first, the presence of non-local terms do not affects symmetry properties. Dismissing integration constants, we have,
\begin{equation}
\mathcal{S} = \frac{2}{{9i\beta }}M^2 A^6 e^{6i(\alpha {\mu } + \beta {\tau })} \left( \begin{array}{l}
 \imath \! \mathbf{i}_2^{}  \\ 
 \sigma _\eta   \\ 
 \end{array} \right) + \int {\mathord{\buildrel{\lower3pt\hbox{$\scriptscriptstyle\smile$}} 
\over r} \partial _\tau  \mathord{\buildrel{\lower3pt\hbox{$\scriptscriptstyle\smile$}} 
\over r} } d\tau 
\end{equation}
and 
\begin{equation}
\mathcal{S}' = \frac{{16}}{{9i\beta }}M^2 A^6 e^{6i(\alpha {\mu } + \beta {\tau })} \left( \begin{array}{l}
 \imath \! \mathbf{i}_2^{}  \\ 
 \sigma _\eta   \\ 
 \end{array} \right) + \int {\mathord{\buildrel{\lower3pt\hbox{$\scriptscriptstyle\smile$}} 
\over r} \partial _\tau  \mathord{\buildrel{\lower3pt\hbox{$\scriptscriptstyle\smile$}} 
\over r} } d\tau . 
\end{equation}
As the amplitude $A$ has in general no dimension, de first term is rigorously a mass term and the mass term difference observed between $\mathcal{S}'$ and $\mathcal{S}$ is said a diffeomorphic mass difference.                                       
Now, this discrete natural discrepancy from diffeomorphic
transformation makes bear in mind as well a visceral type of
discrepancy, the van Dam-Veltman-Zakharov mass-discontinuity for a spin
2 particle. If we consider massive gravitons, uncommon features raise
from field structures in flat space, as verified when Lagrangian
linearized mass term is of Firz-Pauli kind.  Starting from an action
defined by a sum of a cosmological term, the Einstein action and the
spin 2 Fierz-Pauli mass term, 
\\
\begin{center}
\begin{math}\mathcal{L} = \frac{{\sqrt { - \det (g + h)} }}{2}\left[ {R(g + h) -
2\Lambda  + h_{\mu \nu } T^{\mu \nu } } \right] - \frac{{\sqrt { - g}
}}{8}M^2 \left( {h_{\mu \nu } h_{\rho \sigma } g^{\mu \rho } g^{\nu
\sigma } - h_{\mu \nu } h_{\rho \sigma } g^{\mu \nu } g^{\rho \sigma }
} \right)\end{math}\textsf{,}
\end{center}
$$
$$
where \begin{math}g_{\mu \nu } \end{math} is the Einsteinian
metric and \begin{math}h_{\mu \nu } \end{math} is the gravitational
field coupled to an external covariantly conserved source \begin{math}T_{\mu \nu } \end{math}, we gain an unique pole at \begin{math}\nabla ^2  = M^2  - 2\Lambda \end{math} for non-evanescing
\begin{math}\Lambda \end{math} and \begin{math}M\end{math}, with
residue given by, 
$$
$$
\begin{center}
\begin{math}T^{\mu \nu } T_{\mu \nu }  + T\left[ {\frac{{\Lambda  -
M^2 }}{{3M^2  - 2\Lambda }}} \right]T\end{math}\textsf{.}
\end{center}
$$
$$
At fixed $M^2$, when $\Lambda \rightarrow 0$ the residue shrinks to $T^{\mu \nu } T_{\mu \nu }  + \frac{1}{3}T^2 
$ for a massive spin 2 particle. On the other hand, for fixed $\Lambda$, when $M^2 \rightarrow 0$ we find $T^{\mu \nu } T_{\mu \nu }  + \frac{1}{2}T^2$ for a massless spin 2 particle. The van Dam-Veltman-Zakharov mass-discontinuity1is precisely the discrete difference between $T^{\mu \nu } T_{\mu \nu } + \frac{1}{3}T^2 
$ and $T^{\mu \nu } T_{\mu \nu }  + \frac{1}{2}T^2$. Nevertheless, in adS spaces massive and non-massive theories lead to the same predictions. In particular, accordingly Kogan {\it et al.} and Porrati, the limit $M^2 \rightarrow 0$ is smooth in anti-de Sitter space, where  $\Lambda < 0$; from  the  point  of  view  of the supergravity multiplet, this is remarkable and  consistent with the scheme in figure 2, since we may accept a smooth transition  of  the gravitino  field  state $g_\mu$ to its equivalent mirrored graviton field state $G_\nu$ out of the adS  zone within a "reloaded" unbroken SUSY scenario. 
Conversely, still visualizing figure 2, we may take the
inverse way for a massive spin 3/2 particle, assuming fixed
\begin{math}M\end{math} and \begin{math}\Lambda \end{math}
\begin{math}\rightarrow{}\end{math}0 from FLRW space to adS zone
boundary. Although it was prevailing before 1998 the opinion that
graviton is massless, the nonzero graviton mass come being seriously
considered by astrophysical tests for estimation of their limits.  This
supposition is based on the acceptance of the graviton Compton
wavelength as around the Huble distance \\
\begin{center}
\begin{math}c/H\end{math}\begin{math}\approx{}\end{math}\textsf{
1.43X10$^{26}$ \begin{math}m\end{math},}
\end{center}
$$
$$
whence the value of graviton mass, \\

\begin{center}
\begin{math}m_g  \sim \frac{{\hbar H}}{{c^2 }} \approx 1.38 \times
10^{ - 33} eV/c^2 \end{math}\textsf{,}
\end{center}

\begin{center}
\begin{displaymath}
V = \frac{{GM}}{r}\exp ( - r/\lambda _g ),
\end{displaymath}
\end{center}
$$
$$
where \begin{math}M\end{math} is the mass of the source and
\begin{math}\lambda _g \end{math} the Compton wavelength of the
graviton. I will touch upon this subject later, at
the last session of this essay.

The mass of the graviton comes decreasing as the age of the
universe increases. My hypothesis is that the graviton, with its
minimal or even zero mass preserved, heads for the adS boundary across
a declining gradient of \begin{math}\Lambda \end{math}-values at the
adS zone \textquotedblleft{}fuzzy\textquotedblright{} limits. When it
crosses the bounds (\begin{math}\Lambda \end{math}=0), it starts to
acquire gravitino labels.

Now, let us suppose a suitable change in Lagrangian (62), that is,
\begin{equation}
\mathcal{L}_2 = {\rm{ (}}M^2  - 2\Lambda )|A\rangle \langle A^2 |\partial _\tau  \langle A^2 |\int {|A\rangle d\tau  + 1/3 M^2 \langle A^2 |^3  + \mathord{\buildrel{\lower3pt\hbox{$\scriptscriptstyle\smile$}} 
\over r} \partial _\tau  \mathord{\buildrel{\lower3pt\hbox{$\scriptscriptstyle\smile$}} 
\over r} } 
\end{equation}  
where we admit the mass factor in the right first term like the pole presented formerly in the van Dam-Veltman-Zakharov mass-discontinuity. The reason for that is a mass wearing during time evolution. Assuming fixed $M^2$ and \begin{math}\Lambda \rightarrow 0 \end{math} from FLRW space to adS zone boundary, $\mathcal{L}$ turns to expression (62); but, for fixed $\Lambda$, when \begin{math}M^2 \rightarrow 0 \end{math}, $\mathcal{L}$ shrinks to,  
\begin{equation}
\mathcal{L}_3 = {\rm{ }} - 2\Lambda |A\rangle \langle A^2 |\partial _\tau  \langle A^2 |\int {|A\rangle d\tau  + \mathord{\buildrel{\lower3pt\hbox{$\scriptscriptstyle\smile$}} 
\over r} \partial _\tau  \mathord{\buildrel{\lower3pt\hbox{$\scriptscriptstyle\smile$}} 
\over r} } 
\end{equation}
and the mass factor reduces to the double value of the cosmological constant at very large distances from the source in FLRW space. The difference between both Lagrangians is,
\\
\begin{center}
\begin{math} \mathcal{L}_{2-3} = M^2 |A\rangle \langle A^2 |\partial _\tau  \langle A^2 |\int {|A\rangle d\tau  + 1/3 M^2 \langle A^2 |^3  + \mathord{\buildrel{\lower3pt\hbox{$\scriptscriptstyle\smile$}} 
\over r} \partial _\tau  \mathord{\buildrel{\lower3pt\hbox{$\scriptscriptstyle\smile$}} 
\over r} }  - 2\Lambda |A\rangle \langle A^2 |\partial _\tau  \langle A^2 |\int {|A\rangle d\tau }  + 2\Lambda |A\rangle \langle A^2 |\partial _\tau  \langle A^2 |\int {|A\rangle d\tau  - \mathord{\buildrel{\lower3pt\hbox{$\scriptscriptstyle\smile$}} 
\over r} \partial _\tau  \mathord{\buildrel{\lower3pt\hbox{$\scriptscriptstyle\smile$}} 
\over r} } \end{math}\textsf{;}
\end{center}
\begin{equation}
\mathcal{L}_{2-3} = M^2 |A\rangle \langle A^2 |\partial _\tau  \langle A^2 |\int {|A\rangle d\tau  + 1/3 M^2 \langle A^2 |^3 }, 
\end{equation}
where the non-coupled field term was removed.
We must retake equation (62) beneath Euler equation now to derivate the Lagrangian density with respect to the inheritance factor (this is not clear in my former works) defining the time derivative operator, 
\begin{equation}
\left\{ {\langle \mathord{\buildrel{\lower3pt\hbox{$\scriptscriptstyle\smile$}} 
\over G} |,\partial _\tau  \langle \mathord{\buildrel{\lower3pt\hbox{$\scriptscriptstyle\smile$}} 
\over G} |} \right\}_\tau   = \partial _\tau  \langle \mathord{\buildrel{\lower3pt\hbox{$\scriptscriptstyle\smile$}} 
\over G} |\partial _\tau  \langle \mathord{\buildrel{\lower3pt\hbox{$\scriptscriptstyle\smile$}} 
\over G} | + \langle \mathord{\buildrel{\lower3pt\hbox{$\scriptscriptstyle\smile$}} 
\over G} |\partial _\tau ^2 \langle \mathord{\buildrel{\lower3pt\hbox{$\scriptscriptstyle\smile$}} 
\over G} | = 0;
\end{equation}
\begin{equation}
\langle \mathord{\buildrel{\lower3pt\hbox{$\scriptscriptstyle\smile$}} 
\over G} |\partial _\tau ^2 \langle \mathord{\buildrel{\lower3pt\hbox{$\scriptscriptstyle\smile$}} 
\over G} | =  - \partial _\tau  \langle \mathord{\buildrel{\lower3pt\hbox{$\scriptscriptstyle\smile$}} 
\over G} |\partial _\tau  \langle \mathord{\buildrel{\lower3pt\hbox{$\scriptscriptstyle\smile$}} 
\over G} |,
\end{equation}
which may be solved by the following substitutions:
\begin{equation}
p = \partial _\tau  \langle \mathord{\buildrel{\lower3pt\hbox{$\scriptscriptstyle\smile$}} 
\over G} |
\end{equation}
\begin{equation}
p\frac{{dp}}{{d\langle \mathord{\buildrel{\lower3pt\hbox{$\scriptscriptstyle\smile$}} 
\over G} |}} = \partial _\tau ^2 \langle \mathord{\buildrel{\lower3pt\hbox{$\scriptscriptstyle\smile$}} 
\over G} |,
\end{equation}
whence
\begin{equation}
\langle \mathord{\buildrel{\lower3pt\hbox{$\scriptscriptstyle\smile$}} 
\over G} |p\frac{{dp}}{{d\langle \mathord{\buildrel{\lower3pt\hbox{$\scriptscriptstyle\smile$}} 
\over G} |}} + p^2  = 0;
\end{equation}
\begin{equation}
\langle \mathord{\buildrel{\lower3pt\hbox{$\scriptscriptstyle\smile$}} 
\over G} |\frac{{dp}}{{d\langle \mathord{\buildrel{\lower3pt\hbox{$\scriptscriptstyle\smile$}} 
\over G} |}} + p = 0.
\end{equation}
Integral curves are given by,
\begin{equation}
\frac{{dp}}{{d\langle \mathord{\buildrel{\lower3pt\hbox{$\scriptscriptstyle\smile$}} 
\over G} |}} =  - \frac{p}{{\langle \mathord{\buildrel{\lower3pt\hbox{$\scriptscriptstyle\smile$}} 
\over G} |}},
\end{equation} 
with
\begin{equation}
\langle \mathord{\buildrel{\lower3pt\hbox{$\scriptscriptstyle\smile$}} 
\over G} |p = C.
\end{equation} 
This is so because,
\begin{equation}
\frac{{dp}}{{d\langle \mathord{\buildrel{\lower3pt\hbox{$\scriptscriptstyle\smile$}}
\over G} |}} + \frac{p}{{\langle \mathord{\buildrel{\lower3pt\hbox{$\scriptscriptstyle\smile$}} 
\over G} |}} = 0,
\end{equation}
whence,
\begin{equation}
\frac{{dp}}{{d\langle \mathord{\buildrel{\lower3pt\hbox{$\scriptscriptstyle\smile$}} 
\over G} |}} + Pp = Q,P = \frac{1}{{\langle \mathord{\buildrel{\lower3pt\hbox{$\scriptscriptstyle\smile$}} 
\over G} |}},Q = 0,
\end{equation}
$$
$$
and, from $
\rho p = \int {\rho Qd\langle \mathord{\buildrel{\lower3pt\hbox{$\scriptscriptstyle\smile$}} 
\over G} | + C} 
$, we gain
\begin{equation}
\rho  = e^{\int {Pd\langle \mathord{\buildrel{\lower3pt\hbox{$\scriptscriptstyle\smile$}} 
\over G} |} }  = e^{\int {\frac{{d\langle \mathord{\buildrel{\lower3pt\hbox{$\scriptscriptstyle\smile$}} 
\over G} |}}{{\langle \mathord{\buildrel{\lower3pt\hbox{$\scriptscriptstyle\smile$}} 
\over G} |}}} }  = e^{in\langle \mathord{\buildrel{\lower3pt\hbox{$\scriptscriptstyle\smile$}} 
\over G} |}  = \langle \mathord{\buildrel{\lower3pt\hbox{$\scriptscriptstyle\smile$}} 
\over G} |.
\end{equation}
Finally, we have,
\begin{equation} 
\partial _\tau  \langle \mathord{\buildrel{\lower3pt\hbox{$\scriptscriptstyle\smile$}} 
\over G} | = C/\langle \mathord{\buildrel{\lower3pt\hbox{$\scriptscriptstyle\smile$}} 
\over G} |,
\end{equation}
whence,
\begin{equation}
\langle \mathord{\buildrel{\lower3pt\hbox{$\scriptscriptstyle\smile$}} 
\over G} | = C_1 \sqrt \tau .  
\end{equation}
A family of curves from equation (113) is showed in figure 3.
\begin{figure}
\begin{center}
\includegraphics[scale=0.54]{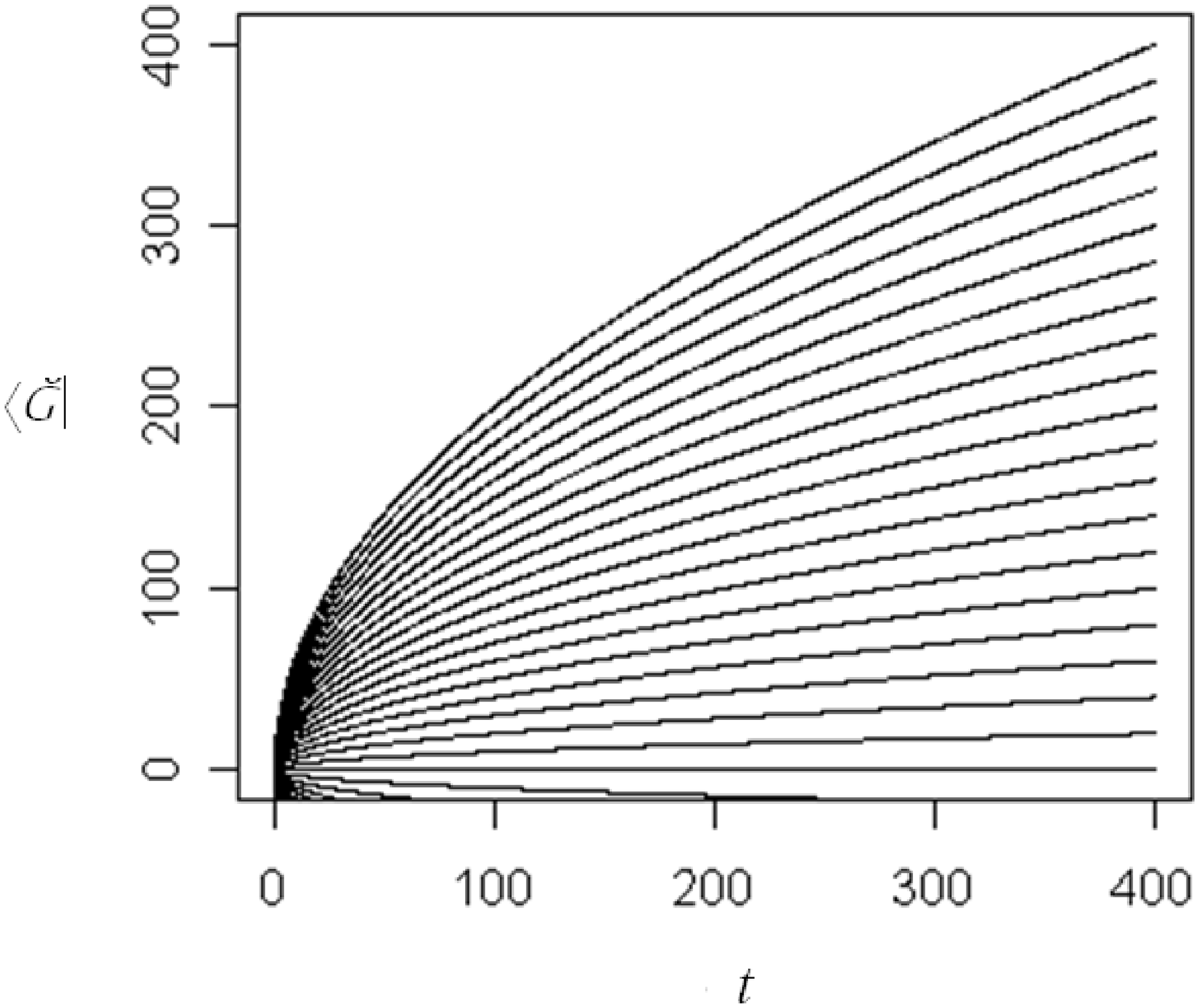}\\ 
\small {{\it Fig. 3- Slice of the field family $\langle \mathord{\buildrel{\lower3pt\hbox{$\scriptscriptstyle\smile$}} 
\over G} | = C_1 \sqrt \tau$ for $C_1$ ranging from 0 to 20.}}
\end{center}
\end{figure}
\\
We conclude that, for present model, if one desires to maintain the spatial side of the field it is stringent to preserve non-local terms in the relations succeeded from the application of canonical procedures. The result would be different if we had a derivative operator as, 
\begin{equation}
\left[ {\langle \mathord{\buildrel{\lower3pt\hbox{$\scriptscriptstyle\smile$}} 
\over G} |,\partial _\tau  \langle \mathord{\buildrel{\lower3pt\hbox{$\scriptscriptstyle\smile$}} 
\over G} |} \right]_\tau   = \partial _\tau  \langle \mathord{\buildrel{\lower3pt\hbox{$\scriptscriptstyle\smile$}} 
\over G} |\partial _\tau  \langle \mathord{\buildrel{\lower3pt\hbox{$\scriptscriptstyle\smile$}} 
\over G} | - \langle \mathord{\buildrel{\lower3pt\hbox{$\scriptscriptstyle\smile$}} 
\over G} |\partial _\tau ^2 \langle \mathord{\buildrel{\lower3pt\hbox{$\scriptscriptstyle\smile$}} 
\over G} | = 0,
\end{equation}
whence,
\begin{equation}
\langle \mathord{\buildrel{\lower3pt\hbox{$\scriptscriptstyle\smile$}} 
\over G} |\partial _\tau ^2 \langle \mathord{\buildrel{\lower3pt\hbox{$\scriptscriptstyle\smile$}} 
\over G} | = \partial _\tau  \langle \mathord{\buildrel{\lower3pt\hbox{$\scriptscriptstyle\smile$}} 
\over G} |\partial _\tau  \langle \mathord{\buildrel{\lower3pt\hbox{$\scriptscriptstyle\smile$}} 
\over G} |,
\end{equation}
whose solution is of type,
\begin{equation}
\langle \mathord{\buildrel{\lower3pt\hbox{$\scriptscriptstyle\smile$}} 
\over G} | = \zeta e^{(\alpha _4 x_4 ...)i}; 
\end{equation}
it would be almost like to get fields with wrong signs, the so-called ghosts of the theory.
Now, we consider operator (102) for distinct fields,
\begin{equation}
\left\{ {\langle g|,\partial _\tau  |\mathord{\buildrel{\lower3pt\hbox{$\scriptscriptstyle\smile$}} 
\over G} \rangle } \right\}_\tau   =  - 2i\langle \mathord{\buildrel{\lower3pt\hbox{$\scriptscriptstyle\smile$}} 
\over G} |\partial _\tau  |g\rangle .
\end{equation}
For $\int {|g\rangle d\tau }$-derivatives, Euler equation furnishes, 
\begin{equation}
\frac{d}{{d\tau }}\left\{ { M^2 |g\rangle {\rm{ }}\langle \mathord{\buildrel{\lower3pt\hbox{$\scriptscriptstyle\smile$}} 
\over G} |\partial _\tau  \langle \mathord{\buildrel{\lower3pt\hbox{$\scriptscriptstyle\smile$}} 
\over G} |} \right\} = 0;
\end{equation}

\begin{equation}
|g\rangle {\rm{ }}\partial _\tau  \langle \mathord{\buildrel{\lower3pt\hbox{$\scriptscriptstyle\smile$}} 
\over G} |\partial _\tau  \langle \mathord{\buildrel{\lower3pt\hbox{$\scriptscriptstyle\smile$}} 
\over G} | + \partial _\tau  |g\rangle \langle \mathord{\buildrel{\lower3pt\hbox{$\scriptscriptstyle\smile$}} 
\over G} |\partial _\tau  \langle \mathord{\buildrel{\lower3pt\hbox{$\scriptscriptstyle\smile$}} 
\over G} | + |g\rangle \langle \mathord{\buildrel{\lower3pt\hbox{$\scriptscriptstyle\smile$}} 
\over G} |\partial _\tau ^2 \langle \mathord{\buildrel{\lower3pt\hbox{$\scriptscriptstyle\smile$}} 
\over G} | = 0;
\end{equation}

\begin{equation}
|g\rangle {\rm{ }}\partial _\tau  \langle \mathord{\buildrel{\lower3pt\hbox{$\scriptscriptstyle\smile$}} 
\over G} |\partial _\tau  \langle \mathord{\buildrel{\lower3pt\hbox{$\scriptscriptstyle\smile$}} 
\over G} | + \langle \mathord{\buildrel{\lower3pt\hbox{$\scriptscriptstyle\smile$}} 
\over G} |\left( {\partial _\tau  |g\rangle \partial _\tau  \langle \mathord{\buildrel{\lower3pt\hbox{$\scriptscriptstyle\smile$}} 
\over G} | + |g\rangle \partial _\tau ^2 \langle \mathord{\buildrel{\lower3pt\hbox{$\scriptscriptstyle\smile$}} 
\over G} |} \right) = 0;
\end{equation}

\begin{equation}
|g\rangle {\rm{ }}\partial _\tau  \langle \mathord{\buildrel{\lower3pt\hbox{$\scriptscriptstyle\smile$}} 
\over G} |\partial _\tau  \langle \mathord{\buildrel{\lower3pt\hbox{$\scriptscriptstyle\smile$}} 
\over G} |  - 2i\partial _\tau  |g\rangle \langle \mathord{\buildrel{\lower3pt\hbox{$\scriptscriptstyle\smile$}} 
\over G} |^2 = 0 ;
\end{equation}

\begin{equation}
|g\rangle {\rm{ }}\partial _\tau  \langle \mathord{\buildrel{\lower3pt\hbox{$\scriptscriptstyle\smile$}} 
\over G} |\partial _\tau  \langle \mathord{\buildrel{\lower3pt\hbox{$\scriptscriptstyle\smile$}} 
\over G} | =  2i\partial _\tau  |g\rangle \langle \mathord{\buildrel{\lower3pt\hbox{$\scriptscriptstyle\smile$}} 
\over G} |^2 .
\end{equation}

This partial differential equation is typed in two complementary parts as,

\begin{equation}
R{\dot \theta } {\dot \theta } =  i2{\theta}^2 {\dot R};
\end{equation}

\begin{equation}
\frac{{\dot \theta } {\dot \theta }} {{\theta}^2} =   i2\frac{{{\dot R}}}{R} = c;
\end{equation}

\begin{equation}
{\dot \theta } {\dot \theta } - c{\theta}^2  = 0;
\end{equation}

\begin{equation}
i2{\dot R} - cR = 0.
\end{equation}

So, we have the following system, 

\begin{equation}
\left\{ \begin{array}{l}
 \partial _\tau  |g\rangle \partial _\tau  \langle \mathord{\buildrel{\lower3pt\hbox{$\scriptscriptstyle\smile$}} 
\over G} | + |g\rangle \partial _\tau ^2 \langle \mathord{\buildrel{\lower3pt\hbox{$\scriptscriptstyle\smile$}} 
\over G} | =  - 2i\partial _\tau  |g\rangle \langle \mathord{\buildrel{\lower3pt\hbox{$\scriptscriptstyle\smile$}} 
\over G} | \\ 
 \partial _\tau  \langle \mathord{\buildrel{\lower3pt\hbox{$\scriptscriptstyle\smile$}} 
\over G} |\partial _\tau  \langle \mathord{\buildrel{\lower3pt\hbox{$\scriptscriptstyle\smile$}} 
\over G} | - c\langle \mathord{\buildrel{\lower3pt\hbox{$\scriptscriptstyle\smile$}} 
\over G} |^2  = 0 \\ 
 i2\partial _\tau  |g\rangle  - c|g\rangle  = 0. \\ 
 \end{array} \right.
\end{equation}
$$
$$ 
Submiting this system to some changes, such as,
\begin{equation}
\partial _\tau  |g\rangle \partial _\tau  \langle \mathord{\buildrel{\lower3pt\hbox{$\scriptscriptstyle\smile$}} 
\over G} | + |g\rangle \partial _\tau ^2 \langle \mathord{\buildrel{\lower3pt\hbox{$\scriptscriptstyle\smile$}} 
\over G} | =  - c|g\rangle \langle \mathord{\buildrel{\lower3pt\hbox{$\scriptscriptstyle\smile$}} 
\over G} |,\nonumber
\end{equation}
\begin{equation}
\partial _\tau  |g\rangle \partial _\tau  \langle \mathord{\buildrel{\lower3pt\hbox{$\scriptscriptstyle\smile$}} 
\over G} | + |g\rangle \partial _\tau ^2 \langle \mathord{\buildrel{\lower3pt\hbox{$\scriptscriptstyle\smile$}} 
\over G} | =  - |g\rangle \frac{{\partial _\tau  \langle \mathord{\buildrel{\lower3pt\hbox{$\scriptscriptstyle\smile$}} 
\over G} |\partial _\tau  \langle \mathord{\buildrel{\lower3pt\hbox{$\scriptscriptstyle\smile$}} 
\over G} |}}{{\langle \mathord{\buildrel{\lower3pt\hbox{$\scriptscriptstyle\smile$}} 
\over G} |}}{\rm{       }}\left( { \div |g\rangle \partial _\tau \langle \mathord{\buildrel{\lower3pt\hbox{$\scriptscriptstyle\smile$}} 
\over G} |} \right),\nonumber
\end{equation}
\begin{equation}
\frac{{\partial _\tau  |g\rangle }}{{|g\rangle }} + \frac{{\partial _\tau ^2 \langle \mathord{\buildrel{\lower3pt\hbox{$\scriptscriptstyle\smile$}} 
\over G} |}}{{\partial _\tau  \langle \mathord{\buildrel{\lower3pt\hbox{$\scriptscriptstyle\smile$}} 
\over G} |}} =  - \frac{{\partial _\tau  \langle \mathord{\buildrel{\lower3pt\hbox{$\scriptscriptstyle\smile$}} 
\over G} |}}{{\langle \mathord{\buildrel{\lower3pt\hbox{$\scriptscriptstyle\smile$}} 
\over G} |}},\nonumber
\end{equation}
\begin{equation}
\frac{{\partial _\tau  |g\rangle }}{{|g\rangle }} =  - \frac{{\partial _\tau ^2 \langle \mathord{\buildrel{\lower3pt\hbox{$\scriptscriptstyle\smile$}} 
\over G} |}}{{\partial _\tau  \langle \mathord{\buildrel{\lower3pt\hbox{$\scriptscriptstyle\smile$}} 
\over G} |}} - \frac{{\partial _\tau  \langle \mathord{\buildrel{\lower3pt\hbox{$\scriptscriptstyle\smile$}} 
\over G} |}}{{\langle \mathord{\buildrel{\lower3pt\hbox{$\scriptscriptstyle\smile$}} 
\over G} |}} = K,\nonumber
\end{equation}
we have
$$
$$
\begin{equation}
\left\{ \begin{array}{l}
 \langle \mathord{\buildrel{\lower3pt\hbox{$\scriptscriptstyle\smile$}} 
\over G} |\partial _\tau ^2 \langle \mathord{\buildrel{\lower3pt\hbox{$\scriptscriptstyle\smile$}} 
\over G} | + \partial _\tau  \langle \mathord{\buildrel{\lower3pt\hbox{$\scriptscriptstyle\smile$}} 
\over G} |\partial _\tau  \langle \mathord{\buildrel{\lower3pt\hbox{$\scriptscriptstyle\smile$}} 
\over G} | + K\langle \mathord{\buildrel{\lower3pt\hbox{$\scriptscriptstyle\smile$}} 
\over G} |\partial _\tau  \langle \mathord{\buildrel{\lower3pt\hbox{$\scriptscriptstyle\smile$}} 
\over G} | = 0 \\ 
 \partial _\tau  |g\rangle  - K|g\rangle  = 0. \\ 
 \end{array} \right.
\end{equation}
$$
$$
The solution of the system and its physical interpretation are matter of the thesis and it is just enough to indicate the equations.
Euler equation referring to ${\langle \mathord{\buildrel{\lower3pt\hbox{$\scriptscriptstyle\smile$}} 
\over G} |}$-derivatives gives,
\begin{equation}
\frac{d}{{d\tau }}\left( {M^2 |g\rangle \partial _\tau  \langle \mathord{\buildrel{\lower3pt\hbox{$\scriptscriptstyle\smile$}} 
\over G} |\int {|g\rangle d\tau  +  M^2 \langle \mathord{\buildrel{\lower3pt\hbox{$\scriptscriptstyle\smile$}} 
\over G} |^2 } } \right) = 0;
\end{equation}
\begin{equation}
|g\rangle ^2 \partial _\tau  \langle \mathord{\buildrel{\lower3pt\hbox{$\scriptscriptstyle\smile$}} 
\over G} | + 2\langle \mathord{\buildrel{\lower3pt\hbox{$\scriptscriptstyle\smile$}} 
\over G} |\partial _\tau  \langle \mathord{\buildrel{\lower3pt\hbox{$\scriptscriptstyle\smile$}} 
\over G} | + \left( {|g\rangle \partial _\tau ^2 \langle \mathord{\buildrel{\lower3pt\hbox{$\scriptscriptstyle\smile$}} 
\over G} | + \partial _\tau  \langle g|\partial _\tau  \langle \mathord{\buildrel{\lower3pt\hbox{$\scriptscriptstyle\smile$}} 
\over G} |} \right)\int {|g\rangle d\tau }  = 0;
\end{equation}
\begin{equation}
\left( {|g\rangle ^2  + 2\langle \mathord{\buildrel{\lower3pt\hbox{$\scriptscriptstyle\smile$}} 
\over G} |} \right)\partial _\tau  \langle \mathord{\buildrel{\lower3pt\hbox{$\scriptscriptstyle\smile$}} 
\over G} | - 2i\partial _\tau  |g\rangle \langle \mathord{\buildrel{\lower3pt\hbox{$\scriptscriptstyle\smile$}} 
\over G} |\int {|g\rangle d\tau }  = 0,
\end{equation}
and we will have a similar system. For all fields we may proceed the same way.
$$
$$
\begin{center}
$\diamondsuit\diamondsuit\diamondsuit$
\end{center}
$$
$$
I have defined some rules which may generate confusion if one
reads without caution. Conventionaly, in a system formed by fermions
and bosons, the measurement operator of conserved charge \begin{math}(
- 1)^F  = \prod\limits_a {\psi _a } \end{math} defines, 

\begin{center}
\begin{math}\left[ {( - 1)^F
,\mathord{\buildrel{\lower3pt\hbox{$\scriptscriptstyle\frown$}}
\over O} _b } \right] = 0\end{math}$_{,}$ for a pure bosonic
operator
\begin{math}\mathord{\buildrel{\lower3pt\hbox{$\scriptscriptstyle\frown$}}
\over O} _b \end{math}$_{,}$ and 
\end{center}

\begin{center}
\begin{math}\left\{ {( - 1)^F
,\mathord{\buildrel{\lower3pt\hbox{$\scriptscriptstyle\frown$}}
\over O} _f } \right\} = 0\end{math}$_{,\textit{ }}$ for a pure
fermionic operator
\begin{math}\mathord{\buildrel{\lower3pt\hbox{$\scriptscriptstyle\frown$}}
\over O} _f \end{math}$_{\textit{ } }$
\end{center}

{\raggedright
(bosonic states have eigenvalue +1; fermionics have 
\textit{\textendash{}}1). Present work settles that,
}

\begin{center}
\begin{math}\left\{ {\langle
\mathord{\buildrel{\lower3pt\hbox{$\scriptscriptstyle\smile$}}
\over G} |,\partial _\tau  \langle
\mathord{\buildrel{\lower3pt\hbox{$\scriptscriptstyle\smile$}}
\over G} |} \right\}_\tau   = 0\end{math} and \begin{math}\left\{ {\langle g|,\partial _\tau  |\mathord{\buildrel{\lower3pt\hbox{$\scriptscriptstyle\smile$}} 
\over G} \rangle } \right\}_\tau   =  - 2i\langle \mathord{\buildrel{\lower3pt\hbox{$\scriptscriptstyle\smile$}} 
\over G} |\partial _\tau  |g\rangle \end{math}.
\end{center}

These brackets have not to say about measurements. They stay
certain physical relations between fields independently of any
observer. The \textit{kets} around the symbols only remember that
fields are subject to gravitorial Wick-rotations.

In this theory, time plays a preponderant role, not only by
non-local considerations: extra dimensions are not necessarily coiled
(it's supposed they wrap back on themselves), but they may exist in a
so little time interval that it's impossible to perceive them.
Furthermore, we'll see that in a theory considering a set of identical
particles time arises like a more intuitive and distinctive physical
background than space, since different positions say little about the
elements of the set.

Equation (60) was interpreted as describing an interaction loop in string
representation (fig. 1); this is an remarkable feature to get in touch
with supergravity standard models, since in a SUSY theory the quadratic
divergences in the scalar Higgs sector are absent because bosonic and
fermionic loop contributions cancel each other. The central idea behind
SUSY as I see it is that an extra-dimension is in fact defined by the
presence of a singularity on a configuration $\mathbb{R}^n$-manifold, that is, an
identified pair of point-states (if we can get from one to the other by
a rotation); the same particle appearing as a boson or a fermion, one
species emerging as two. When identified, they lose their labels (mass,
charge, spin, savor, or color). In my opinion, the only rotation able
to identify physically two point-states is an imaginary rotation, or a
Wick-rotation. Let \begin{math}\vartheta \end{math} be a subspace of $\mathbb{R}^n$; each point
in \begin{math}\vartheta \end{math}\textbf{ }is glued together with a
point-partner by \textbf{O}(\textit{k,l:k+l=n}) Wick-rotations (global
isometries). The quotient space \begin{math}\mathbb{R}^n/\vartheta \end{math} is said a
representation orbifold $\bar O$. This is a rational way to distinct
physical process from pure mathematic manipulation. In mathematics,
nothing substantial is claimed to identify two geometric points on a
manifold; it's sufficient to declare it formally. In physics, such a
thing requires a payment in energy and a change of microphysical
status: a particle that fuses with your
\textquotedblleft{}shadow\textquotedblright{} counterpart transforms
itself in a particle without any labels, and a change that
\textquotedblleft{}maps\textquotedblright{} a boson into a fermion, or
makes a fermion emerge from a boson, must be an imaginary
transformation. The orbifold $\bar O$ represents a new manifold, the
adS zone, with 4 singular glued locus, the co-dimensions of the theory
or generators of its symmetries.
 We can see the pair of inseparable particles as an object and
its image. They are not exactly identical, since an image is not equal
to its material source, and they rest at different places. But, let us
suppose they can interchange their mutual condition of object and
image, that is, the image is transformed in the object and
\textit{vice-versa}, according to certain natural rules. An elegant
manner to represent this situation is to convert the object, by a
suitable mathematic rule of transformation, in its image or imaginary
partner; at the same time, we transmute the image in the source object
by the inverse application. I called the transformation \begin{math}g
= \Upsilon G\end{math}(or \begin{math}G = \Upsilon ^ - 
g\end{math}) an enantiomorphism. We can understand
this dual process considering identical particles, one graviton letting
back this condition and changed to a gravitino, and another graviton
just getting this status, letting back gravitino labels.  

Taking only the affinors,  
it remains for gravitons, 

\begin{center}
\begin{math}\langle
\mathord{\buildrel{\lower3pt\hbox{$\scriptscriptstyle\smile$}}
\over G} _\mu  |^2 = \end{math}\begin{math}\left\{ {\left[
\begin{array}{l}
\; \; \; \mathds{1}_2^{}  \\
- \mathds{1}_2^{}  \\
\end{array} \right],\left[ \begin{array}{l}
\; \; \; \mathds{1}_2^{}  \\
- \mathds{1}_2^{}  \\
\end{array} \right],\left[ \begin{array}{l}
\; \; \; \mathds{1}_2^{}  \\
- \mathds{1}_2^{}  \\
\end{array} \right],\left[ \begin{array}{l}
\mathds{1}_2^{}  \\
\mathds{1}_2^{}  \\
\end{array} \right]} \right\}\end{math}\textsf{.}
\end{center}
This implies graviton's quadratic signature in gravitorial
coordinates as expressed by \begin{math}({}_ - ^ + \mathds{1},{}_ - ^ + \mathds{1},{}_ - ^ + 
\mathds{1},{}_ + ^ +  \mathds{1})\end{math}, with longitudinal (bottom) signs reminding
the Minkowskian form.
 For gravitinos, we write the equivalent quadratics,
\begin{center}
\begin{math}\langle
\mathord{\buildrel{\lower3pt\hbox{$\scriptscriptstyle\smile$}}
\over G} _\nu  |^2 = \end{math}\begin{math}\left\{ {\left[
\begin{array}{l}
- \mathds{1}_2^{}  \\
\; \; \; \mathds{1}_2^{}  \\
\end{array} \right],\left[ \begin{array}{l}
- \mathds{1}_2^{}  \\
\; \; \; \mathds{1}_2^{}  \\
\end{array} \right],\left[ \begin{array}{l}
- \mathds{1}_2^{}  \\
\; \; \; \mathds{1}_2^{}  \\
\end{array} \right],\left[ \begin{array}{l}
- \mathds{1}_2^{}  \\
- \mathds{1}_2^{}  \\
\end{array} \right]} \right\},\end{math}
\end{center}
which provides a symmetric signature \begin{math}({}_ + ^ -  \mathds{1},{}_ +
^ -  \mathds{1},{}_ + ^ -  \mathds{1},{}_ - ^ -  \mathds{1})\end{math}. Because its strong
symmetry, expression (60) is said to be the \textbf{equation of mirroring}.

Considering full fields on, we see that our supersymmetric
transformations are clearly related with spacetime ones. For a
gravitational source \begin{math}G_{kj} \end{math}, we have,
\begin{equation}
h_{ki}  \approx \mathord{\buildrel{\lower3pt\hbox{$\scriptscriptstyle\smile$}} 
\over G} _{ki}  = G_{ki}  - \frac{1}{2}\delta _{ki} G_{\eta \eta }. 
\end{equation}
\subsection*{Field dual lattices}
\addcontentsline{toc}{subsection}{Field dual lattices}
It would be useful to take two distinct field functions for a bidimensional cell construction, creating a background over which we may evaluate another field; so, we describe the virtual broadcast of the latter using the former two field functions as purveyors of a plane coordinate system. Of course, there is a natural generalization to higher dimensions, but that is not the point here. More precisely, we are dealing with a two-dimensional Voronoi tessellation, an exhaustive unique plane partition in convex cells; each cell is formed by a single centroidal nucleus and a set of points closer to that nucleus more than to any other.
We want to know what value of gravitino field is expected given two graviton field functions furnishing flat coordinates. This is very interesting as a forequantum approach, since we can do further probabilistic interpretations looking at the cell construction (each cell as a volume of probability). One by one, graviton dihedral dual volumes contribute with portions of total gravitino field intensity.  
Let us take the independent solutions \begin{math}\langle
\mathord{\buildrel{\lower3pt\hbox{$\scriptscriptstyle\smile$}}
\over G} |_1  = C_1 \sqrt \tau  \end{math} and \begin{math}\langle
\mathord{\buildrel{\lower3pt\hbox{$\scriptscriptstyle\smile$}}
\over G} |_2  = \frac{1}{4}c\tau ^2 \end{math} as variables of generic
functions \begin{math}\xi _1  = f_1 (\langle
\mathord{\buildrel{\lower3pt\hbox{$\scriptscriptstyle\smile$}}
\over G} |_1 )\end{math} and \begin{math}\xi _2  = f_2 (\langle
\mathord{\buildrel{\lower3pt\hbox{$\scriptscriptstyle\smile$}}
\over G} |_2 )\end{math}. The nucleus of a complete Voronoi cell is a centroid defined as, 
\begin{equation}
C_j  = \frac{{\int_A {x\rho (x)dA} }}{{\int_A {\rho (x)dA} }},
\end{equation}
\begin{figure}[h]
\begin{center}
\includegraphics[scale=0.64]{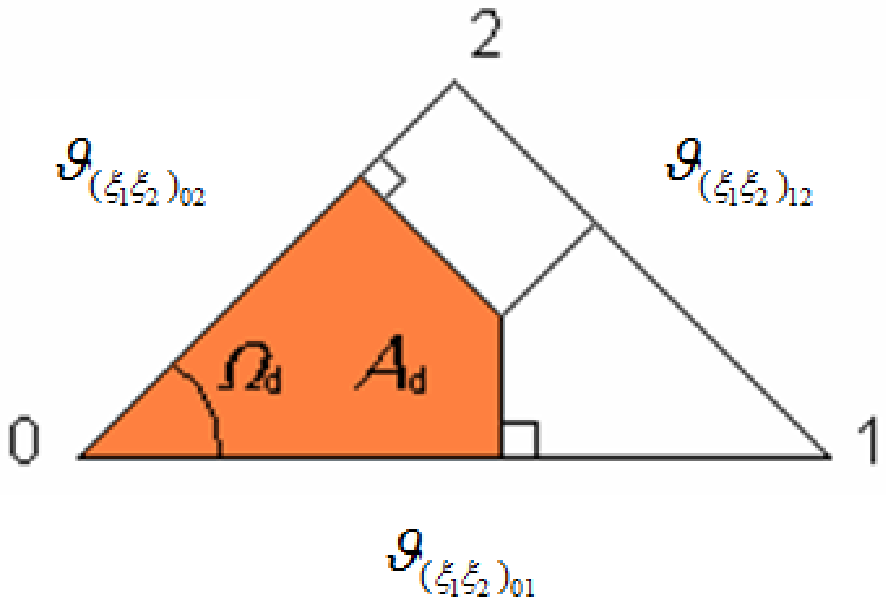}\\ 
\small {{\it Fig. 4- Dual area $A_d$ from vertex $0$ within the angle ${\Omega}_d$.}}
\end{center}
\end{figure}
where $A$ is the area of the cell and $\rho(x)$ the density function. In a more tangible physical sense, if the density is constant the centroid is equivalent to the centre of mass. A centroidal Voronoi diagram may be seen as a chart that represents a lower-limit energy state, since it renders minimal the integral,
\begin{equation}
\int_A {\rho (x)\left| {C_j  - x} \right|^2 }. 
\end{equation}
Finally, the dihedral field volume contribution for the vertex 0 in figure 4 is given by,
\begin{equation}
A_d \left( {\vartheta ^2 } \right) = \frac{1}{{16A}}\left[ {\vartheta ^2 _{(\xi _1 \xi _2 )_{12} } (\vartheta ^2 _{(\xi _1 \xi _2 )_{01} }  + \vartheta ^2 _{(\xi _1 \xi _2 )_{02} } ) - (\vartheta ^2 _{(\xi _1 \xi _2 )_{01} }  - \vartheta ^2 _{(\xi _1 \xi _2 )_{02} } )^2 } \right].
\end{equation}
\begin{figure}[h]
\begin{center}
\includegraphics[scale=0.54]{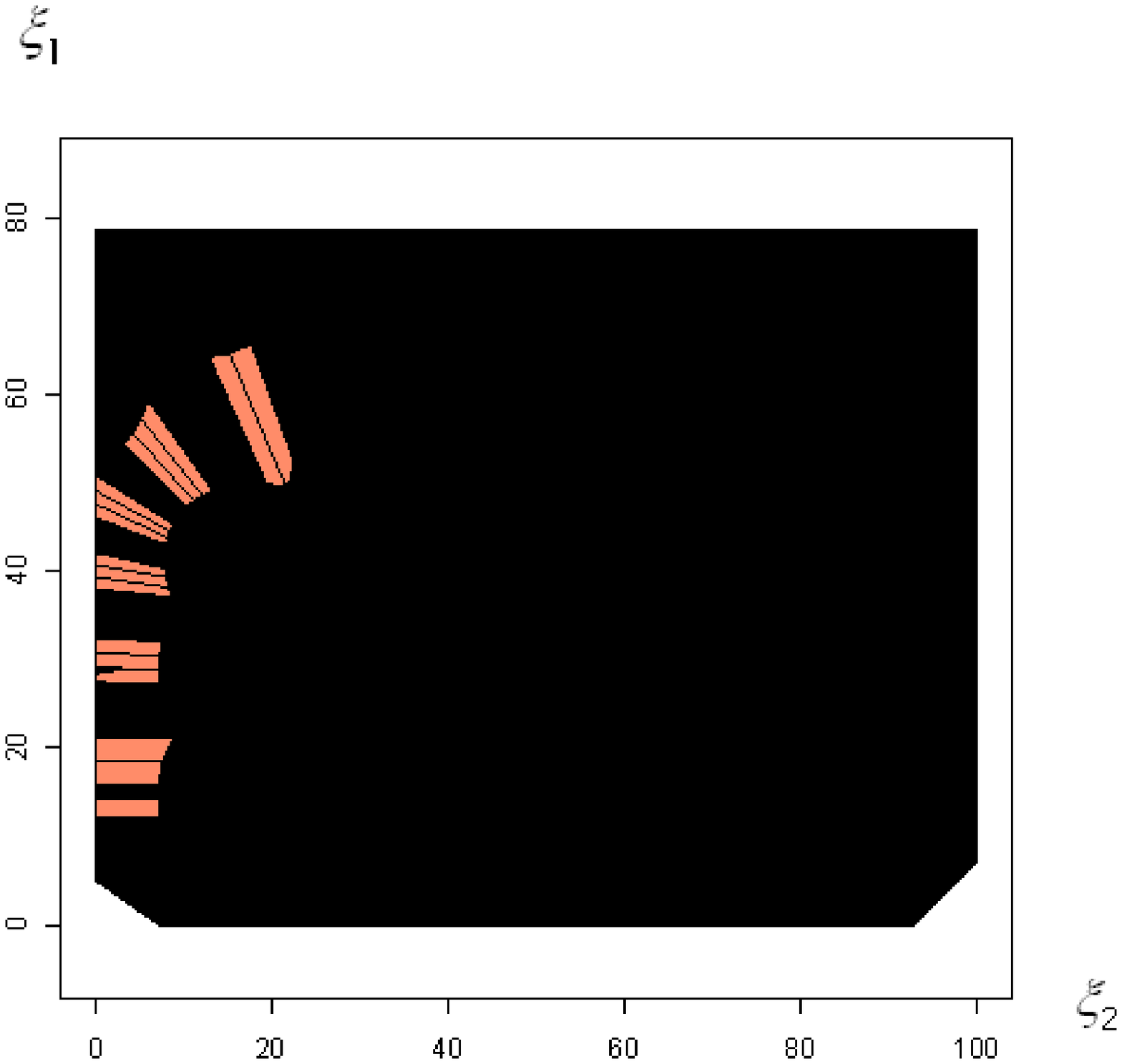}\\ 
\small {{\it Fig. 5- Voronoi tessellation about quantiles from a periodic field solution (salmon cells upon the dark background) accordingly. The background may be seen as a vacuum and the colourful ribbons as its excitations.}}
\end{center}
\end{figure}
In a practical sense, these tessellations form combined mosaics of different field solutions, included those not truly "on shell" solutions; it's a manner to represent how that fields cohabit and how they may create a ground to detach some special excitation. Periodic field solutions in figure 5 evolve by "wave fronts" in a parabolic way within a certain interval of coordinates $\xi_1$ and $\xi_2$.
It is worth nothing that this approach is a purely heuristic procedure that enable us to identify geometrical patterns of correlation or general casualistic associations that may clarify system's physics. In fact, figure 5 offers a possible cut off for supergravity, since global supersymmetry warrants a lower bound for the vacuum energy, say $\left\langle 0 \right|H\left| 0 \right\rangle  \ge 0$.\\
\begin{center}
$$
$$
$\diamondsuit\diamondsuit\diamondsuit$
\end{center}
$$
$$
$$
$$

\newpage

\thispagestyle{empty}
\mbox{} 
\vfill
\begin{center}
\begin{flushleft}
\begin{caixa4}
{\LARGE Chapter 3}
\end{caixa4}
\end{flushleft}
\end{center}
\vfill
\vfill
\pagebreak

\section*{ A brief note on strings }
\addcontentsline{toc}{section}{A brief note on strings}
$$
$$
\hrule
\markright{\bfseries A brief note on strings} 
\vspace{4ex}
However I'm not an enthusiast of the string model nowadays, and I recognize that some explanations on string theory are beyond my understanding, an
ideal of mine in this subject is to see a theory that tells about a plausible
and credible world, understandable without impositions by hand,
absolutely clear in four dimensions, or at least with few intelligible
additional well cleared up dimensions. At this moment, unfortunately,
there is no place to this dream at all. As Warren Siegel said few years
ago about what can we expect from string theories,
\textquotedblleft{} \ldots{}known string theories have little
or no predictive power. They are thus less a 'Theory of
Everything' than a 'Theory of
Nothing'$\;$\textquotedblright{} (Siegel, 2001). Today, nobody even know the full equations of the theory and how the Standard Model in all its details would follow from it. In addition, there is a large number of {\it aprioristic} possible string models. Meanwhile, I have some words to say.
\subsection*{Strings and gravity in common approach} 
\addcontentsline{toc}{subsection}{Strings and gravity in common approach}
With respect to gravity, every consistent string theory must contain a massless spin-2
state, whose interactions reduce at low energy to general relativity. The common effective string theory equations governing the low-energy dynamics of the gravitational field and of its sources are not postulated {\it ad hoc}, but they are required for the consistency of a quantum theory of strings propagating in a curved manifold, and interacting with other fields possibly present in the
background. Usually, the gravitational massless-bosonic sector of the string effective action contains
the metric and, at least, one more fundamental field, a scalar field $\phi$, the so-called "dilaton". The corresponding action (known as "tree-level action") is written as,
\begin{equation}
S =  - \frac{1}{{2\lambda _\mathtt{S}^{\mathtt{d} - 1} }}\int {d^{\mathtt{d} + 1} } x\sqrt {\left| g \right|} e^{ - \phi } \left[ {R + \left( {\nabla \phi } \right)^2 } \right] + S_\Sigma   + S_m, \nonumber
\end{equation}
where $\left( {\nabla \phi } \right)^2  = \nabla _\mu  \phi \nabla ^\mu  \phi$. Here $S_\Sigma$ is the boundary term required to gain Einstein's field equations in the general-relativistic limit, and $S_m$ is the action of the other fields, supposedly coupled to $\phi$ and to $g_{\mu\nu}$ accordingly the conformal invariance of fundamental string interactions. The above equation is written adopting the so-called "string frame" (S-frame)\footnote{The S-frame is the frame in which it holds the uniqueness of the fixed coupling to a constant dilaton for all fields at the tree-level. The S-frame is, in an aesthetic physical sense, the "favorite" string theory frame, since in such frame physical intuition is more easy to apply.} parameterization of the action with $\phi$ dimensionless. The metric $g_{\mu\nu}$ is the same metric to which a fundamental string is minimally coupled, and
with respect to which a free "test" string evolves geodesically. The constant length $\lambda _\mathtt{S}$ represents the characteristic proper extension of a quantized one-dimensional object like a fundamental
string, providing the natural units of length and energy $(\lambda _\mathtt{S}^{ - 1}  = M_\mathtt{S})$ for a physical model based on the S-frame action. The independent equation governing the dynamics of the dilaton field is obtained by varying the above defined action with respect to $\phi$.
\subsection*{Strings and inflation}
\addcontentsline{toc}{subsection}{Strings and inflation} 
Models of inflation should be constructed in the context of supergravity. We might consider the scalar fields in supergravity as coordinates of a Kähler manifold. The metric on this manifold is given by ${K_{\bar \phi \phi } }$, where the Kähler potential $K\left( {\phi ,\bar \phi } \right)$ is a real function of $\phi$ and its hermitian conjugate ${\bar \phi }$. It is worthwhile to assume an effective action during inflation of the form,
\begin{equation}
S = \int {d^4 x\sqrt { - g} } \left[ { - \frac{1}{2}R + g^{\mu \nu } \partial _\mu  \bar \phi K_{\bar \phi \phi } \partial _\nu  \phi  - V\left( {\phi ,\bar \phi } \right)} \right], \nonumber
\end{equation}
using the flat Robertson-Walker ansatz,
\begin{equation}
ds^2  = dt^2  - a(t)^2 dX^2, \; \phi  = \phi (t), \nonumber 
\end{equation}
where ${\partial _\mu  \bar \phi K_{\bar \phi \phi } \partial _\nu  \phi }$
resumes the scalar kinetic terms.
 For this flat model, we consider the Hubble parameter $H$ defined as $H \equiv \dot a/a$. Inflation requires $- \dot H/H^2  \ll 1$, that is, $3H^2  \simeq V$. Thus, the energy density of the Universe should be dominated by the scalar potential. The slow-roll equations of motion,
\begin{equation}
\dot \phi  =  - 3HK_{\bar \phi \phi }^{ - 1} V_{\bar \phi },\nonumber 
\end{equation}
hold as an approach for the dynamics of the scalar fields, and we might assume that they have been attained for all epochs of interest. Here, we define the canonically normalised inflaton $\sigma$ by,
\begin{equation}
\frac{1}{2}d\sigma ^2  = d\bar \phi K_{\bar \phi \phi } d\phi .\nonumber   
\end{equation}
For inflation, the necessary conditions may be expressed in terms of the potential
with, 
\begin{equation}
\left( {\frac{{V'}}{V}} \right)^2  \ll 1, \; \left| {\frac{{V''}}{V}} \right| \ll 1, \nonumber   
\end{equation}
where a prime denotes the derivative with respect to $\sigma$.
\subsection*{Strings in gravitorial approach}
\addcontentsline{toc}{subsection}{Strings in gravitorial approach}  
 Now, we must effort ourselves to get a classic picture of strings
before any trial to build superstrings. Figure 1 shows a scheme that combines gravitors and strings. Let us take this way,
considering present theory. Given the contour, 
\begin{center}
\begin{math}\langle
\mathord{\buildrel{\lower3pt\hbox{$\scriptscriptstyle\smile$}}
\over G} |_X  = \langle
\mathord{\buildrel{\lower3pt\hbox{$\scriptscriptstyle\smile$}}
\over G} |_{X + 2\pi } ,{\rm{ }}|g\rangle _X  = |g\rangle _{X + 2\pi }
\end{math}\textit{$_{  }$},
\end{center}
it leads, for the Lagrangian density which describes
field diffusion of graviton and gravitino,
\begin{center}
\begin{math}T\int {d^2 X{\rm{ }}\mathcal{L}\left( {\partial _\tau  \langle
\mathord{\buildrel{\lower3pt\hbox{$\scriptscriptstyle\smile$}}
\over G} |,{\rm{ }}\int {|g\rangle d\tau ,{\rm{ }}\partial _\tau 
|g\rangle ,{\rm{ }}\int {\langle
\mathord{\buildrel{\lower3pt\hbox{$\scriptscriptstyle\smile$}}
\over G} |d\tau ,{\rm{ }}\langle
\mathord{\buildrel{\lower3pt\hbox{$\scriptscriptstyle\smile$}}
\over G} |{\rm{, }}|g\rangle } } } \right){\rm{ }}} \eta  =
S\end{math}\begin{math}\end{math};
\end{center}

\begin{center}
\begin{math}T\int {d^2 X{\rm{ }}\left[ { - 4\left( {\partial _\tau 
\langle \mathord{\buildrel{\lower3pt\hbox{$\scriptscriptstyle\smile$}}
\over G} |\int {|g\rangle d\tau  + \partial _\tau  |g\rangle \int
{\langle \mathord{\buildrel{\lower3pt\hbox{$\scriptscriptstyle\smile$}}
\over G} |d\tau } } } \right) - \Upsilon \langle
\mathord{\buildrel{\lower3pt\hbox{$\scriptscriptstyle\smile$}}
\over G} |^2  - 2\langle
\mathord{\buildrel{\lower3pt\hbox{$\scriptscriptstyle\smile$}}
\over G} ||g\rangle  - \Upsilon ^ -  |g\rangle ^2 } \right]{\rm{
}}\eta  =} \end{math}
\end{center}
\begin{equation}
= S.\nonumber
\end{equation}
Here, $\eta$ is a special metric related to the
vacuum refringence. Gravitors bring habitual intrinsic metric, not as
independent variable like the brane world-volume metric
$h^{\alpha \beta}$ in the covariant action of Polyakov,
\\
\begin{center}
\begin{math}S =  - \frac{T}{2}\int {d^{p + 1} \xi \sqrt { - \det h}
\left\{ {h^{\alpha \beta } \partial _\alpha  X^\mu  \partial _\beta 
X_\mu   - (p - 1)} \right\}} \end{math},
\end{center}
\vspace{0.2in}
but inside the polarization itself. A new metric $\eta$ was
introduced, adding the refringence of the medium. The idea is that
fields react by the refringence, so revealing deformations that
correspond to the gravitational effects. Of course, I based the
study on the simplest case of string propagation in adS$_{3}$ space at
the non-quantum mechanical limit as described by Maldacena and
Ooguri\footnote{ J. Maldacena \& H. Ooguri, \textquotedblleft{}Strings
in adS$_{3}$ and $SL(2,R)$ WZW Model: I\textquotedblright{},
hep-th/0001053.}. So, it seems natural to retake the subject from some fundamental
ideas.
We may say that the earliest string theory was in fact first introduced by
d'Alembert in 1747. It is very interesting to look at the small
fragment of a string, crudely speaking, \textit{a la} \textit{cuerdas
vibrantes} (fig. 3.1), showing that the force of oscillation resulting
from the acting tension supposed constant all along in the string is,
\\
\begin{center}
\begin{math}F = T\left( {\partial _{x^1 }^2 \langle
\mathord{\buildrel{\lower3pt\hbox{$\scriptscriptstyle\smile$}}
\over G} |} \right)dx^1 \end{math},
\end{center}
\vspace{0.2in}
the well-known expression of the vibrating strings;
\begin{math}{\large F}\end{math} acts on the external medium
\textbf{iff} it stands in harmonic mode of graviton, an object with
force status.Eventually, as we know, a strong signal that new physics should appear
at very high energy is that the union of gravity with quantum theory yields a nonrenormalizable
quantum field theory. As models of unified theories, modern superstrings brought serious implications to our understanding about the early stages of the universe evolution and many formal inconveniencies when we deal with scales below the Planck scale by means of a four dimensional low energy effective field theory derived from them.

\begin{figure}
\begin{center}
\includegraphics[scale=0.44]{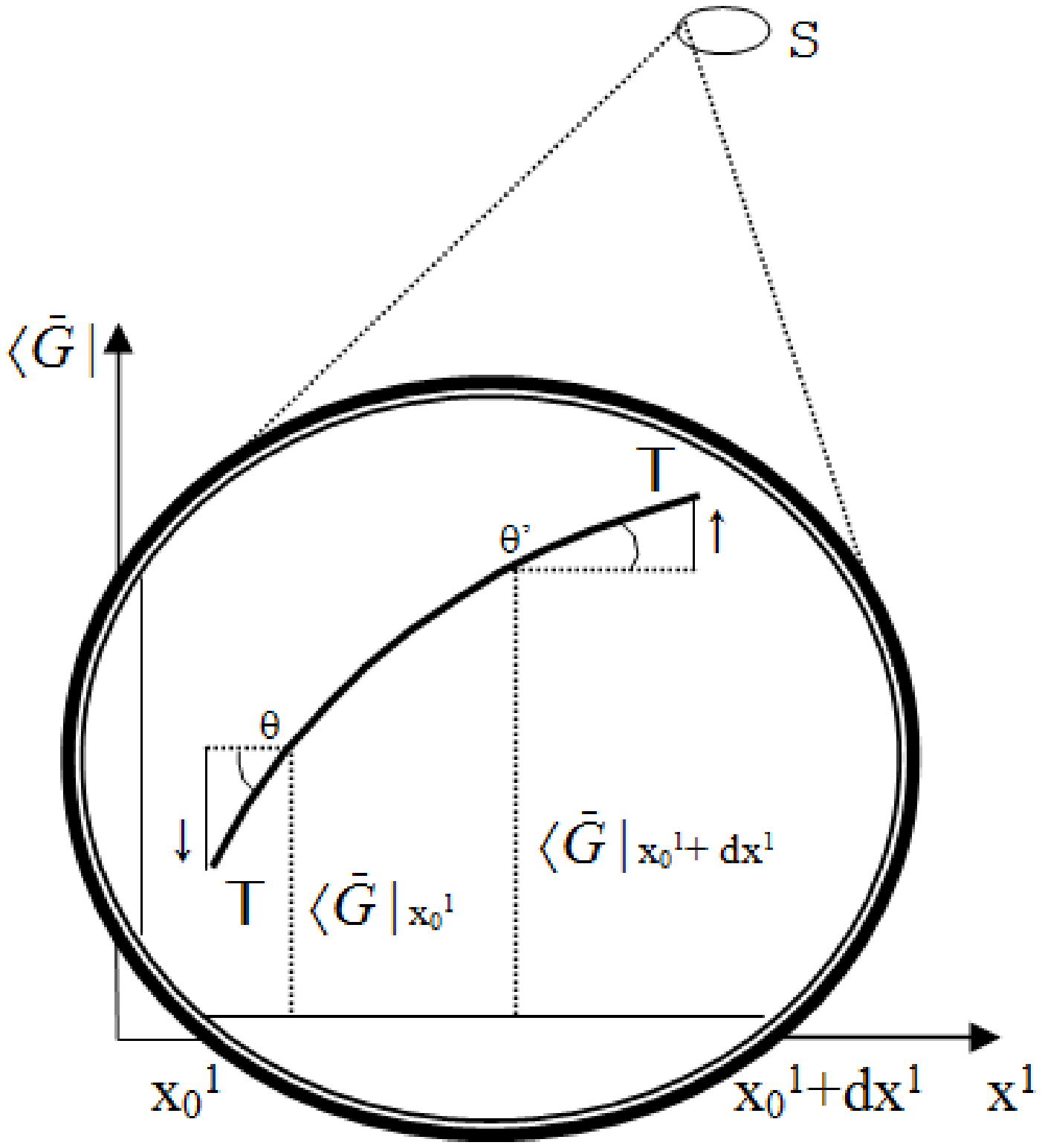}\\
\small {{\it Fig. 6- A zoom on a piece of superstring and detailing of the field action.}}
\end{center}
\end{figure}
\begin{center}
$$
$$
$\diamondsuit\diamondsuit\diamondsuit$
\end{center}

\newpage

\thispagestyle{empty}
\mbox{} 
\vfill
\begin{center}
\begin{flushleft}
\begin{caixa5}
{\LARGE Chapter 4}
\end{caixa5}
\end{flushleft}
\end{center}
\vfill
\vfill
\pagebreak

\section*{The SCYL program ({\bf Supersymmetric Cosmology at Yonder Locals})}
\addcontentsline{toc}{section}{The SCYL program ({\bf Supersymmetric Cosmology at Yonder Locals})} 
$$
$$
\hrule
\begin{flushright} 
\small {\it The typical lifetime of a new trend in high-energy physics
and cosmology nowadays is about 5 to 10 years. If it survived for a longer time, the chances are 
that it  will be with us for quite a while.}
\\
\vspace{3.2mm}
$\mathpzc{Andrei \hspace{0.4mm} Linde}$
\end{flushright}
\markright{\bfseries The SCYL program} 
\vspace{4ex}
At small length scales it was observed deviations from the postulated homogeneity of the Universe at large scales, a fact that imposes 1) - the need to investigate whether the accelerated cosmological expansion is real, that is, whether the acceleration is not an effect of the inhomogeneity, and 2) - the necessity to look for the length scale from which the Universe becomes homogeneous, if indeed it is. 
Among several inhomogeneous cosmological models, the Lemaître-Tolman (LT) model - the simplest and, in my opinion, the only practicable in fact - have been applied with some interesting results as an alternative to explain the universe without cosmological constant at scales ${\cal O}(10)h^{ - 1} Mpc$ or even larger. The LT metric under the assumption of spherical symmetry in simultaneously synchronous and commoving frame can be read as a branch of solutions of the equation,
\begin{equation}
ds^2  =  - dt^2  + b^2 \left( {r,t} \right)dr^2  + R\left( {r,t} \right)^2 \left( {d\theta ^2  + \sin ^2 \theta {\rm{ }}d\phi ^2 } \right)
\end{equation}
that describes an inhomogeneous collapse of dust or, which comes to be the same, its time reversal. These solutions are given by, 
\begin{equation}
b^2  = \frac{{R'\left( {r,t} \right)^2 }}{{1 + f\left( r \right)}},
\end{equation}
where the function $f(r)$ can be thought as a spatial curvature parameter and is one of the three classical LT arbitrary functions, and $R$ is the angular diameter distance.
In spite of the challenges it faces and the objections faced to its major presuppositions, which one expects from a LT model is its simultaneous and reasonable agreement with data from cosmic microwave background (CMB), from type Ia supernova, from structure formation and so forth. For example, Alnes {\it et al}. (2006) showed that a LT region which reduces to an Einstein-de Sitter cosmology at a radius of $1.4 Gpc$ can match both the supernova data and the location of the first acoustic peak in the CMB.
The SCYL program is an attempt to build a model of Universe that, at the same time, treats the problem of the threshold of inhomogeneity from which the Universe becomes perturbatively near a homogeneous and isotropic Friedmann-Robertson-Walker model and the hypothesis of a supersymmetric cosmology based on a unified view of cosmology and supergravity in terms of a meta-field theory, as described formerly. 
\subsection*{The heavens in G-decay}
\addcontentsline{toc}{subsection}{The heavens in G-decay} 
It is possible, although not yet confirmed, that stellar systems are assailed by a cumulative gravitational effect to which I named G-decay (G of Gravitational). In fact, the word "decay" was applied here to make reference to the decomposition in fragments of an exploding planet. The idea incorporates in a certain manner the notion of a gravitational field in continuous regeneration due to LeSage \footnote{G.L. LeSage, Berlim Mem. 404, 1784.}, the prime physicist to establish a gravitational model of particles flow. Being massive bodies condensations of the gravitational continuum, in this model they would be absorbing constantly gravitic energy. At normal conditions, the thermodynamical equilibrium is preserved while the body carries on a cooling gradient proportional to the heat gained from the gravitons flow. Instabilities in planet's kernel, meanwhile, may start perturbative occurrences in the gradient, making the planet become hot with the impact of gravitons. Of course, this situation leads to an anomalous state of excessive energy with repercussion all over the planetary structure until the creation of lethal fractures. It is interesting to note that continuous regeneration of the field may be due to the permanent feedback of gravitons from gravitinos and conversely.
Another problem that thrives among cosmologists and astrophysicists is the nature of the active core of objects like giant elliptical galaxy NGC 4261. Since black-holes were accepted as the powerful kernel of many galaxies, new puzzles are tantalizing scientific community. In conformity to ideas of Hoyle and Narlikar, a massive black hole in galaxy nuclei may attain enough density to provoke inversion of the matter dragging accordingly its accretion disk. Energy discharges would outburst until a new contraction of the core begins the converging flow. Intermittent ejections of matter observed in the nucleous of several galaxies would be well understood by emitions and absorptions of gravitons, alternating into the adS zone as the fueling and refueling processes of the central accretion disk. Why? What factors lead to an outcome of galaxy evolution and do not to another? Why certain galaxies haven not black-hole kernels (observations of the spiral M33 seem to indicate that it does not enclose a black-hole)? 
I have not all the answers, but let us consider dense objects having an anti-de Sitter (adS) "shell" within which supergravity plays a fundamental role, a role that may shed new lights on the evaporation of light Primordial Black Holes \footnote{In supergravity models, the evaporation of light Primordial Black Holes
(PBHs) should be a source of gravitinos.}. Thus, a black-hole has its adS shell, whose boundary defines the so-called "event horizon" \footnote{The adS spacetime arises as ground state in supergravity and does not represent our living world. Precisely by this reason, it seems to be a suitable background for the physics behind the limits of an event horizon.}. Theoretically, bodies not so strong also have adS shells; they serve to shelter interactions that occur when a body crosses a clump of dark matter, since it is necessary to explain facts that we never access directly. The filtrino field sets a gateway to a gravitino dominated world; if one prefers, the gateway would be a type of portal that breaks supersymmetry. In fact, the famous "event horizon", viewed only within the supergravity approach, begins at the "filtrinosphere" of the source. Transitions across this layer are best brought before the mind as reloading field states.
To understand the mechanism of gravitational cataclysms, we must to turn back to figure 2. Physical orbifold ${\bar O}$ drafts the "appearance" of a process that involves the decay of spinless filtrino ${\mathord{\buildrel{\lower3pt\hbox{$\scriptscriptstyle\smile$}} 
\over r} }$ - when touched by external graviton - to gravitino $g$ via enantiomorphic transformation $S(\mathds{1}, g)$. Inspired in works of Hamaguchi {\it et al.}, I adopt some premises with empirical content; my assumption is that, being gravitino the lightest supersymmetric particle, scalar filtrino plays the role of next-to-lightest supersymmetric long-lived particle with main decay mode,
\begin{equation}
\mathord{\buildrel{\lower3pt\hbox{$\scriptscriptstyle\smile$}} 
\over r}  \to rg_{3/2}, 
\end{equation} 
and lifetime estimated as,
\begin{equation}
\mathcal{T}_{\mathord{\buildrel{\lower3pt\hbox{$\scriptscriptstyle\smile$}} 
\over r} }^{ - 1}  \simeq 9days \times (M_{3/2} /10Gev)^2  \times (150Gev/M_{\mathord{\buildrel{\lower3pt\hbox{$\scriptscriptstyle\smile$}} 
\over r} } )^5 . 
\end{equation}
Gravitino mass can be deduced cinematically from,
\begin{equation}
M_{3/2}^2  = M_{\mathord{\buildrel{\lower3pt\hbox{$\scriptscriptstyle\smile$}} 
\over r} }^2  + M_r^2  - 2M_{\mathord{\buildrel{\lower3pt\hbox{$\scriptscriptstyle\smile$}} 
\over r} } E_r, 
\end{equation}
whence,
\begin{equation}
{\rm{\gamma }}_\eta ^ -  |g\rangle ^2  = {\rm{\gamma }}_\eta  \frac{{ - (M_2  - \mathord{\buildrel{\lower3pt\hbox{$\scriptscriptstyle\smile$}} 
\over \varepsilon } )}}{{\sqrt {M_{\mathord{\buildrel{\lower3pt\hbox{$\scriptscriptstyle\smile$}} 
\over r} }^2  + M_r^2  - 2M_{\mathord{\buildrel{\lower3pt\hbox{$\scriptscriptstyle\smile$}} 
\over r} } E_r } }}\langle \mathord{\buildrel{\lower3pt\hbox{$\scriptscriptstyle\smile$}} 
\over G} |^2, 
\end{equation}
where $E_r$ is the energy of the emitted filtrino. This last equation talks about enantiomorphism between graviton and gravitino, exposing filtrino mediation.
According to previous discussion, filtrinosphere would imply a surmountable wall behind which gravitinos exist in appreciable quantity; it is possible an abnormal great number of gravitinos colliding with filtrinos, reheating adS zone and scattering a lot of gravitons to FLRW space. The reason for this anomaly was unclear until recently, but now I think that, from the view point of gravity, it is related to 1) a general collapse tendency of massive bodies, each of them with a distinct final outcome according to its initial mass, and so, according to its orbifold limits (adS radius), or to 2) a dive of the body into a dark matter dense region. As Afsar Abbas said,\\
\\
"The passage of the earth through dense clumps of dark matter would produce large quantities of heat in the interior of this planet through the capture and subsequent annihilation of dark matter particles. This heat would lead to large-scale volcanism which could in turn have caused mass extinctions. The periodicity of such volcanic outbursts agrees with the frequency of palaeontological mass extinctions as well as the observed periodicity in the occurrence of the largest flood basalt provinces on the globe. Binary character of these extinctions is another unique aspect of this signature of dark matter. In addition dark matter annihilations appear to be a new source of heat in the planetary systems"\footnote{A. Abbas, arXiv:astro-ph/9910265 v1 14 Oct 1999.}.
\\\\
Conventional Big-Bang model pictures just post-inflation universe as a thermal gas of particles and superparticles from which gravitinos were generated by collisions. Gravitinos, so relics of that period, might take part in the core of dense heaven bodies and now are submitted to the action of colliders always the core is embedded into a clump of dark matter. In this scenario, heat generation results from all annihilation channels which in some way create photons, including gravitino-filtrino inelastic collisions. From works of Press and Spergel, we may say this heat is generally given by, 
\begin{equation}
\mathop Q\limits^ \bullet{_E}  = e\mathop {N_E }\limits^ \bullet  m_x, 
\end{equation}
where $\mathop {N_E }\limits^ \bullet$ is the capture rate of WINPs for the body, $e$ is the annihilation quantity for heat production and $m_x$ is the mass of the dark matter particle. 
The equilibrium of the transformations held by filtrinos tells much about the source behavior and it is the focus of the present proposal. For the discussion of such mechanisms, the assumption of gravitons with mass not equal to zero might bring important consequences. That being said, since a positive cosmological constant makes the expansion accelerate, the study of $\Lambda$ inside adS zone and its implications for the speeding-up universe takes a fundamental part in the study. Because of proposed theory of gravitino adS confinement, free gravitinos - those flowing across the immensities - are much less numerous than gravitons.

Let us go back to reading the pictures. I defined an orbifold $\bar O$ as a quocient space, 
\[\mathbb{R}^n/\vartheta.
\]
This orbifold is a representation of the adS zone. But an anti-de Sitter space is a Lorentzian manifold with negative constant sectional curvature. Suppose a hiperbolic space,
\[
{\rm{H}}^n  = \left( {\mathbb{R}^+   \times S^{n - 1} ,\sinh ^{- 2} \rho \left( {d\rho ^2  + d\sigma {}_0^2 } \right)} \right),
\]
with $d\sigma {}_0^2$ as the standard metric on $S^{n - 1}$. By the way, the topology of adS$_{n+1}$ is $S^1 \times 
\mathbb{R}^n$ (timelike circle represents the compact dimension). The $n+1$ dimensional adS space $\mathbb{S}$ for, 
\[
ds^2  =  - \coth ^2 \rho d\tau ^2  + \sinh ^{ - 2} \rho \left( {d\rho ^2  + d\sigma {}_0^2 } \right),
\]
may be seen also as the warp product between $\mathbb{R}$ and $\rm{H}^n$, and, taking into account subspace $\vartheta$, $\mathbb{S}= \left( {\mathbb{R} \times {\rm{H}}^n /\vartheta, ds^2} \right)$, similar to the known vacuum solution of Einstein's field equation for negative cosmological constant, $\mathbb{V}= \left( {\mathbb{R} \times {\rm{H}}^n, ds^2 } \right)$. The adS space $\mathbb{S}$ is a hyperbolic cylinder with four gaps (singularities). Of course, time slices $\rm{H}^n$ have not geometrical relations with figure 2, since that picture is only a metaphorical scheme associated to a dense spherical body and not a configuration space. All that I assume is the following: very near the event horizon, the background geometry is approximated by the geometry of an adS spacetime.  

\subsection*{The caves of the hiddens}
\addcontentsline{toc}{subsection}{The caves of the hiddens}
With supergravity on one hand, at least in theory, gravitinos are thermally produced in the early Universe by QCD (Quantum Chromodynamical) processes. If they exist, in conventional field theory they are fermions of spin $3/2$ obeying Rarita-Schwinger equation. In particular, there are advantages in studying gravitino's properties when its mass is in the superlight range because it can be easily emitted in astrophysical processes such as supernova cooling and neutron star cooling. If gravitinos are stable, and have the mass $10 - 100 \hspace {2 mm} GeV$, the relic density associated will be cosmologically relevant and gravitinos can play the role of the cold dark matter.
 With cosmology on the other hand, however the standard analysis of light propagation settles that the Universe is perturbatively almost homogeneous and isotropic, observations show a foam-like structure, with clusters of galaxies stitched by long filaments and bubbles surrounding huge voids. Friedmann-Robertson-Walker (FRW) model  is manifestly not true on scales smaller than $70-100h^{-1}Mpc$. Besides, more recent surveys suggest that $\sim 45\%$ of the volume of the universe is occupied by voids of a characteristic scale $30h^{-1} Mpc$, where $h$ is the dimensionless Hubble parameter, $H_0 = 100h \hspace {2 mm}km \hspace {2 mm} sec^{-1}Mpc^{-1}$. An attempt to describe the observed voids of matter is the assumption of a LT Swiss-cheese model, in which the inhomogeneous spherical cavities are enclosed by a friedmannian background. Some cavities, containing an adS massive central core, are isolated by a filtrinosphere border (see figure 2). 
The final goal is to explain supergravity into the LT spacetime and try to examine supersymmetric gravitino production at the cavities having dense cores in order to identify some particular pattern correlated to the limit-wall beyond which the universe becomes near homogeneous. The introduction of such Swiss-cheese model has the particular interest to impose a physical constraint to the gravitino's influence (the decay law $1/r^3$). The LT void would be the region of major probability to find gravitinos. It is important to remember that gravitino is only one candidate to compose the total amount of dark matter. It is also remarkable that voids without central dense cores may contain gravitinos as relics of early stages. 
\subsection*{On the implementation of extra-dimensions}
\addcontentsline{toc}{subsection}{On the implementation of extra-dimensions}
The introduction of extra-dimensions in the frame of a theory may be seem, at first, an arbitrary act, mainly by the fact that such dimensions are unseen, that is, they are hidden and not so readily imaginable. The very complex mathematical tools required in supergravity and string theory, with ten or eleven dimensions, allow many physicists skeptical and even averse about the physical content of such theories. As Dirac disciples, we still believe in natural laws depicted by a mathematics so simple as elegant. Thereby, from an epistemological point of view, to explain the reasons of hidden extra-dimensions sounds artificial and {\it ad hoc}. But, since Kaluza and Klein observed that electromagnetism can be considered as part of a five-dimensional gravity theory, where the fifth dimension is curled up, modern physics have edited several theories with such extra-dimensions.
An alternative to try to explain why extra-dimensions are hidden is to postulate they are very small and curved on themselves, {\it i. é}, they are compactfied \footnote{The extra dimensions would describe a compact space of sufficiently small size, so that they can only be probed by very high energies.}. Other possibility, not-compact, was proposed by Paul Wesson, of the Canadian University of Waterloo. In his cosmological model, he argues that we live in an Universe of five dimensions; the fifth dimension would be the geometry manifested as substance when observed by sentient beings that live in a hypersurface of four dimensions. Lisa Randall, of the University of Princeton, and Raman Sundrum, of the University of Boston, also proposed a model in which we live on a spacetime hypersurface of five dimensions, now called the "Randall-Sundrum model". Randall and Sundrum assume our four dimensional universe as the boundary of a five dimensional "brane". In such model, the graviton string can move into the fifth dimension, while the other bosonic strings cannot. In both models, the authors preserved all mathematic formalism developed by Einstein in his general relativity, only changing dimensionality to five.    
Personally, I understand that such extra-dimensions explain in fact the entailed symmetries. It is not the case to observe measurable dimensions as width and height, but to verify, directly or not, the reality of such symmetries by empirical methods. The symmetries are special hierarchly higher dimensions required for all that we may observe in the ordinary four dimensions. I would like to quote an excerpt of my Master Thesis about the introduction of a fifth dimension in the LT cosmology:
\\
\\
{\it "Recently, however, I was interested in the possibility to describing a Lemaître-Tolman cavity by means of a 5D metric embedded in a standard friedmannian background in 4D. The idea was to assume that inhomogeneity carries in the fifth dimension an information able to give it a symmetry such that its structure remains irreducible to FLRW, unless on the junction between the LT cavity and the FLRW background. Here goes a brief discussion about cosmological symmetries from a geometric point of view. The symmetries of a spacetime, or isometries, constitute a group in which a) the identity is an isometry, b) the inverse of an isometry is an isometry, and c) the composition of two isometries is an isometry. We define the orbit of a point $p$ as the set of all points for which $p$ can be moved by the action of the translative isometries of space. The orbits are necessarily homogeneous, that is, all physical quantities are exactly the same at each point. Once an invariant variety is a set of points mappable in themselves by the group of isometries, the orbits are necessarily invariant manifolds. The freedom of translation in a given space, or transferring dimension, is generally symbolized by the letter "$\rm{s}$", being defined the expression $\rm{s}\le \rm{n}$, where $n$ is the number of dimensions of the space.

An important  subgroup of the group of isometries, whose size can be seen at each $p$, is the isotropy group, that is, the group of isometries which leave $p$ fixed (rotations). In general, the rotational dimension of a given space is represented by the letter "$\rm{q}$", being defined the expression ${\rm{q}} \le {\rm{1/2 n(n - 1)}}$, where $n$ is the number of dimensions of the space. Thus, the dimension $\mathfrak{D}$ of the isometry group of that space 
is $\mathfrak{D} = \rm{s} + \rm{q}$ (translations + rotations). Indeed, the continuous isometries 
are generated by the Lie algebra of Killing vectors. The action group is characterized 
by the nature of its orbit in the space. For a cosmological model, due to the four-dimensionality of the spacetime, the possible orbital dimensionalities are $\rm{s} = 0, 1, 2, 3, 4$. The group of isometries which characterizes the LT models in 4D is the $\mathcal{G}_{\rm{s+q}} = \mathcal{G}_{\rm{3}}$ or $\mathcal{G}(\rm{2, 1})$, isomorphic to the real special pseudo-orthogonal group in ${\rm{s+q}}$, $\mathcal{SO}(\rm{2}, \rm{1})$. Each LT model is 
characterized by a two-dimensional surface with spherical symmetry: $\rm{s} = 2$; all observations anywhere on the surface are rotationally symmetric around a privileged space direction: $\rm{q} = 1$; so, $\mathfrak{D}_{LT4D} = \rm{2} + \rm{1} = \rm{3}$. 

However, the implementation of a fifth angular dimension corresponds to the introduction of an extra "translational" degree of freedom, $\rm{p} = 1$, whence $\mathfrak{D}_{LT5D} = \rm{s} + \rm{p} + \rm{q} = \rm{2} + \rm{1} + \rm{1} = \rm{4}$. Therefore, the LT model in 5D, as stated earlier, requires a group of isometries $\mathcal{\bar G}_{\rm{4}}$, isomorphic to the singular orthogonal group in $\rm{s+p+q}$, $\mathcal{SO}(\rm{2}, \rm{1}, \rm{1})$, corresponding to the inhomogeneous Lie algebra $\rm{so}(\rm{2}, \rm{1}, \rm{1})$. I concluded that, outside the junction, the only way to get an LT metric in 4D reducible to FLRW would be by an unknown mechanism of symmetry spontaneous broken $\parallel\mathcal{Q}\parallel$, such that $\parallel\mathcal{Q}\parallel\mathcal{SO}(\rm{2}, \rm{1}, \rm{1})\longrightarrow \mathcal{G}_{\rm{3}}$\footnote{The unique situation that is physically and clearly need to find a metric LT in 4D reducible to FLRW occurs in the junction, where $\Sigma$ has to be four-dimensional.}. This study, still underway, has been discussed with researchers in field theory and is based on the belief that the universe evolves conserving material symmetry between homogeneous and inhomogeneous regions, being this symmetry  only broken by the breaking mechanism above mentioned"}.

\subsection*{The present stage of the SCYL program}
\addcontentsline{toc}{subsection}{The present stage of the SCYL program}
The SCYL program is an interdisciplinary project involving field theory, relativistic cosmology, astrophysics and philosophy of science. In particular, it treats the subject of how human mind constructs cosmological knowledge. The main steps of the SCYL program are:
\begin{enumerate}
\item Overview of the classical/quantum field theories - {\bf 80\% concluded};
\item Introduction to meta-field theory focusing symmetry between the superpartners of gravity (graviton and gravitino) - {\bf  80\% concluded};
\item Overview of the Lemaître-Tolman inhomogeneous cosmology (with $\Lambda = 0$) - {\bf 100\% concluded};
\item Investigation of the current theories on the structure of the Universe (mainly those referring to the dark matter composition, formation of voids, etc.) with philosophical criticism - {\bf  to perform};
\item Formulation of a cosmology based on a symmetry between homogeneous and inhomogeneous regions - {\bf  60\% concluded};
\item Formulation of a model to explain voids as Lemaître-Tolman bubbles of dark matter; most of this matter is allegedly composed by relic cold gravitinos (bubbles without dense cores) or by young hot gravitinos (bubbles with dense cores) - {\bf 40\% concluded};
\item Search for patterns of distribution of voids so that we may infer, based on the expected gravitino presence, the threshold from which the Universe may be considered closely homogeneous - {\bf to perform};
\item Search for an unified understanding of the cosmological constant - {\bf to perform}.
\item Concluding remarks about theoretical results - {\bf to perform}.  
\end{enumerate}
The program will be terminated in two and one-half years, from $2011$ to $2014$.
\begin{center}
$$
$$
$\diamondsuit\diamondsuit\diamondsuit$
\end{center}

\newpage

\thispagestyle{empty}
\mbox{} 
\vfill
\begin{center}
\begin{flushleft}
\begin{caixa6}
{\LARGE Exercises}
\end{caixa6}
\end{flushleft}
\end{center}
\vfill
\vfill
\pagebreak

\section*{Exercises}
\addcontentsline{toc}{section}{Exercises}
\pagestyle{myheadings}
\markright{\bfseries Exercises} 
\vspace{4ex}
1)- Given the lagrangian,
\begin{equation}
\mathcal{L}_2 = M^2 |g\rangle \langle \mathord{\buildrel{\lower3pt\hbox{$\scriptscriptstyle\smile$}} 
\over G} |\partial _\tau  \langle \mathord{\buildrel{\lower3pt\hbox{$\scriptscriptstyle\smile$}} 
\over G} |\int {|g\rangle d\tau }  + 1/3 M^2 \langle \mathord{\buildrel{\lower3pt\hbox{$\scriptscriptstyle\smile$}} 
\over G} |^3  + \mathord{\buildrel{\lower3pt\hbox{$\scriptscriptstyle\smile$}} 
\over r} \partial _\tau  \mathord{\buildrel{\lower3pt\hbox{$\scriptscriptstyle\smile$}} 
\over r},\nonumber  
\end{equation} 
and the current,    
\begin{equation}
\frac{{\partial \mathcal{L}}}{{\partial \partial _4 |g\rangle }}\partial _4 |g\rangle  - \mathcal{L} = j,\nonumber 
\end{equation}
define the auxiliary field ${\mathord{\buildrel{\lower3pt\hbox{$\scriptscriptstyle\smile$}} 
\over r} }$ so that Noether theorem for classic fields holds as we assume, 
\begin{equation}
\langle \mathord{\buildrel{\lower3pt\hbox{$\scriptscriptstyle\smile$}} 
\over G} | = \frac{{ - Ae^{ - (1 + i) \tau  } }}{{\sqrt 2 }} + K_1.\nonumber  
\end{equation} 
{\bf Hint}: apply Euler equation to $\mathcal{L}_2$ for $\partial _4 \langle \mathord{\buildrel{\lower3pt\hbox{$\scriptscriptstyle\smile$}} 
\over G} |$ and calculate prime version of ${\mathord{\buildrel{\lower3pt\hbox{$\scriptscriptstyle\smile$}} 
\over r} }$; apply $\langle \mathord{\buildrel{\lower3pt\hbox{$\scriptscriptstyle\smile$}} 
\over G} |$ and ${\mathord{\buildrel{\lower3pt\hbox{$\scriptscriptstyle\smile$}} 
\over r} }$ in current equation above and adjust constants.\\\\
2)- Proof that lagrangian,
\begin{equation}
\mathcal{L}_1  =  - 4\left( {\partial _\tau  \left| G \right\rangle \int {\left| g \right\rangle d\tau  + \partial _\tau  \left| g \right\rangle \int {\left| g \right\rangle d\tau } } } \right) - \Upsilon \left| G \right\rangle ^2 - \nonumber  
\end{equation}
\begin{equation}
- 2\left| G \right\rangle \left| g \right\rangle  - \Upsilon ^ -  \left| g \right\rangle ^2,\nonumber  
\end{equation}
has symmetry $O(n)$ with respect to field derivatives $\partial _\tau  |g\rangle$ and $\partial _\tau  \langle \mathord{\buildrel{\lower3pt\hbox{$\scriptscriptstyle\smile$}} 
\over G} |$.
\\
\\ 
\\
{\bf Hint}: apply Euler equation to $\mathcal{L}_1$; compare results with classical $O(2)$ infinitesimal transformations   $\delta q_1  = \alpha q_2$ and $\delta q_2  =  - \alpha q_1$; confront invertibility of $\mathbb{C}_{k,l}$ - elements in found expressions and symmetry between $\delta q_1$ and $\delta q_2$.\\\\
3)- Let suppose we assume gravitorial non-Abelian regime given by,
\begin{equation}
g_{\mu \nu }  =  - g_{\nu \mu }. \nonumber 
\end{equation}
Proof that the commutator of two parametrized transformations, 
\begin{equation}
\left[ {\delta |g\rangle ,\delta (\partial _\tau  \langle \mathord{\buildrel{\lower3pt\hbox{$\scriptscriptstyle\smile$}} 
\over G} |)} \right] =  \pm 2\delta |\varpi \rangle \nonumber  
\end{equation}
and 
\begin{equation}
\varpi  = f\left( {|g\rangle ,\partial _\tau  \langle \mathord{\buildrel{\lower3pt\hbox{$\scriptscriptstyle\smile$}} 
\over G} |} \right),\nonumber  
\end{equation}
is a transformation itself with parameter $\varpi$ being an analytical function of $|g\rangle$ and $\partial _\tau  \langle \mathord{\buildrel{\lower3pt\hbox{$\scriptscriptstyle\smile$}} 
\over G} |$. Show that this fact configures symmetry.
\\
\\
\\
{\bf Hint}: apply
\begin{equation}
\partial _\tau  \langle \mathord{\buildrel{\lower3pt\hbox{$\scriptscriptstyle\smile$}} 
\over G} | = \left[ \begin{array}{l}
 \mathds{1}_2  \\ 
 \sigma _1  \\ 
 \end{array} \right]\left( {\frac{{A(1 + i)e^{ - (1 + i) {\tau }} }}{{\sqrt 2 }}} \right) \nonumber
\end{equation}
and
\begin{equation}
|g\rangle  = \left[ \begin{array}{l}
  \; \; \imath \! \mathbf{i}_2  \\ 
 {\scriptscriptstyle{-}}\sigma _2  \\ 
 \end{array} \right]Ae^{ - (1 + i) {\tau }}. \nonumber
\end{equation}
4)- Write the gravitational plane wave considering the gravitor of polarization and deduce a Lagrangian density for the interaction of the wave with a particle starting from the Minkowskian metric $d\tau ^2  =  - \eta _{\mu \nu } dx^\mu  dx^\nu$. \\
\begin{center}
$\diamondsuit\diamondsuit\diamondsuit$
\end{center}

\newpage
\thispagestyle{empty}
\mbox{} 
\vfill
\begin{center}
\begin{flushleft}
\begin{caixa3}
{\LARGE Articles and books}
\end{caixa3}
\end{flushleft}
\end{center}
\vfill
\vfill
\pagebreak

\markright{\bfseries Articles and books} 

$$
$$
\begin{center}
$\diamondsuit\diamondsuit\diamondsuit$
\end{center}
\end{document}